\documentclass[useAMS,usenatbib,usegraphicx]{mn2e}

\def\degK{\,{}^\circ\mbox{K}}

\def\kpc{\mbox{\,kpc}}
\def\pc{\mbox{\,pc}}

\def\Gyr{\mbox{\,Gyr}}
\def\muG{\,\mu\mbox{G}}
\def\kms{\mbox{\,km s}^{-1}}

\def\pcc{\mbox{\,cm}^{-3}}

\title[Gaseous and Stellar Orbits]{Comparing Gaseous and
Stellar Orbits in a Spiral Potential}

\author[G\'omez, Pichardo \& Martos] {Gilberto C. G\'omez$^{1}$ \thanks{E-mail:
g.gomez@crya.unam.mx},
B\'arbara Pichardo$^2$ \thanks{E-mail: barbara@astro.unam.mx} and
Marco A. Martos$^{2}$ \thanks{E-mail: marco@astro.unam.mx}
\\
$^{1}$Centro de Radioastronom\'\i a y Astrof\'\i sica,
Universidad Nacional Aut\'onoma de M\'exico, Apdo. Postal 3-72,
Morelia Mich. 58089, M\'exico\\
$^2$Instituto de Astronom\'ia,
Universidad Nacional Aut\'onoma de M\'exico,
Apdo. Postal 70-264, Ciudad Universitaria D.F. 04510, M\'exico}

\date{Accepted for publication in MNRAS}

\begin{document}

\maketitle

\label{firstpage}

\begin{abstract}

It is generally assumed that gas in a galactic disk follows closely
non self-intersecting periodic stellar orbits. In order to test this
common assumption, we have performed MHD simulations of a
galactic-like disk under the influence of a spiral galactic
potential. We also have calculated the actual orbit of a gas parcel
and compared it to stable periodic stellar orbits in the same
galactic potential and position. We found that the gaseous orbits
approach periodic stellar orbits far from the major orbital
resonances only.
Gas orbits initialized at a
given galactocentric distance but at different azimuths can be
different, and scattering is conspicuous at certain galactocentric
radii. Also, in contrast to the stellar behaviour, near the 4:1 (or
higher order) resonance the gas follows nearly circular orbits, with
much shorter radial excursions than the stars. Also, since the gas
does not settle into a steady state, the gaseous orbits do
not necessarily close on themselves.

\end{abstract}
 

\section{Introduction}
\label{sec:intro}

Differences between stellar orbital dynamics and gas dynamics in disk
galaxies play a fundamental role in the interpretation (or
misinterpretation) of kinematics observations in order to
deduce general physical
properties of galaxies.
One example \citep{gom06} is that the inferred quantity of dark matter
in galaxies can be affected by inconsistencies between the well-studied
gas rotation curve and the relatively poorly studied stellar disc
rotation curve.

Several observational and theoretical studies have been devoted to the
investigation of the relationship between the gas streamlines
and the corresponding stellar orbits. In particular, it is customary
to identify the non self-intersecting stable periodic orbits of a given
astronomical
system, with plausible regions for gas streams to settle
(e.g. \citealt{kal73, sim73, lin74, wie74, kru76, san76, van78,
alb82, san83, con89, ath92a, ath92b, pat94, pin95, eng97, pat97, eng00,
veg01, reg03, pat09}).

From resonances emerge strong orbital families like the x1 type that
manage to survive in models up to approximately the 4:1
resonance.
After that periodic orbits from many families are found.
This means that at resonances new
families of periodic orbits start existing. At resonances we may
encounter more than one family of periodic orbits
\citep{con86,hel96,kau05}.
This is in part intrinsic to the problem, due to the non-linear
character of the perturbations.
In normal non-barred spiral galaxies with a
moderate non-axisymmetric component, such as spiral arms, the
stellar orbits and gaseous flow can be highly non-linear (this
means, the response of a dynamical system to a perturbation depends
non-linearly on the amplitude of the perturbation), even far from
resonances. Also, phenomena like gas shocks or stellar chaos
are found even in the case of low-amplitude spiral perturbations
\citep{pic03,per12}.

Large scale galactic gaseous and stellar flows present various
important differences in their kinematics.
Some interesting known differences between the response of
gas and stars to an imposed galactic potential model are due to the
fact that gas is very responsive and suffers violent shocks, unlike
stars. In resonant regions, it has been suggested that
shocks occur close to the intersections of stable periodic
orbits
\citep{sch81, alb82, ath92b}.
Gas is attracted to the spiral arms by their gravitational
field and, unlike stars, it responds to
pressure gradients.  In non-barred galaxies with large bulges or thick
disks (such as early spirals), stars are supported not only by rotation
but by strong radial velocities or velocity dispersion,
while gas settles down in much
thinner disks supported by rotation. The later the morphological
type of the galaxy, the more
similar are gas and stellar kinematics \citep{bec04}.
In this work we study the interplay between gas flow and stellar orbital
dynamical properties of spiral galaxies.
We have calculated a set of MHD
and particle simulations
in a static background gravitational potential based
on a full 3D spiral galactic model \citep{pic03}.
This includes spiral arms different in nature to the classic
tight winding approximation,
in order to contribute to the understanding of
the relationship between stellar orbits and gas orbits in non-barred
spiral galaxies, and the relation with periodic orbits inside and outside
resonance regions.

This paper is organized as follows.
In \S \ref{sec:model}, the
galactic potential used to compute the MHD and stellar orbits and the
initial simulation set up is described.
In \S \ref{sec:orbits}, the stellar and MHD
orbits obtained with the spiral potential are presented and compared
with each other.
In \S \ref{sec:conclusions}, we present a brief summary and
discussion of our results.

\begin{figure}
  \includegraphics[width=0.85\hsize]{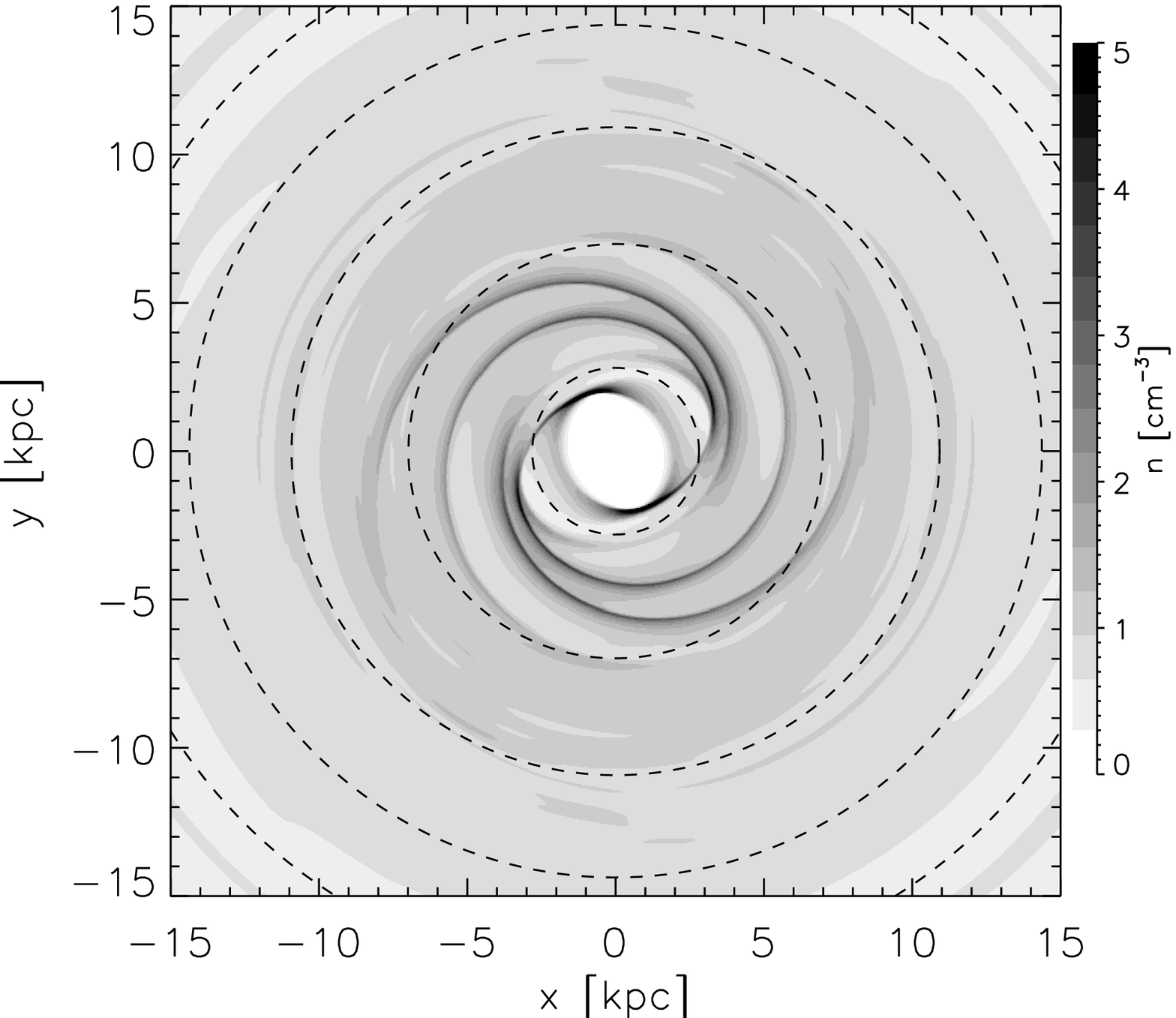}\\
  \includegraphics[width=0.85\hsize]{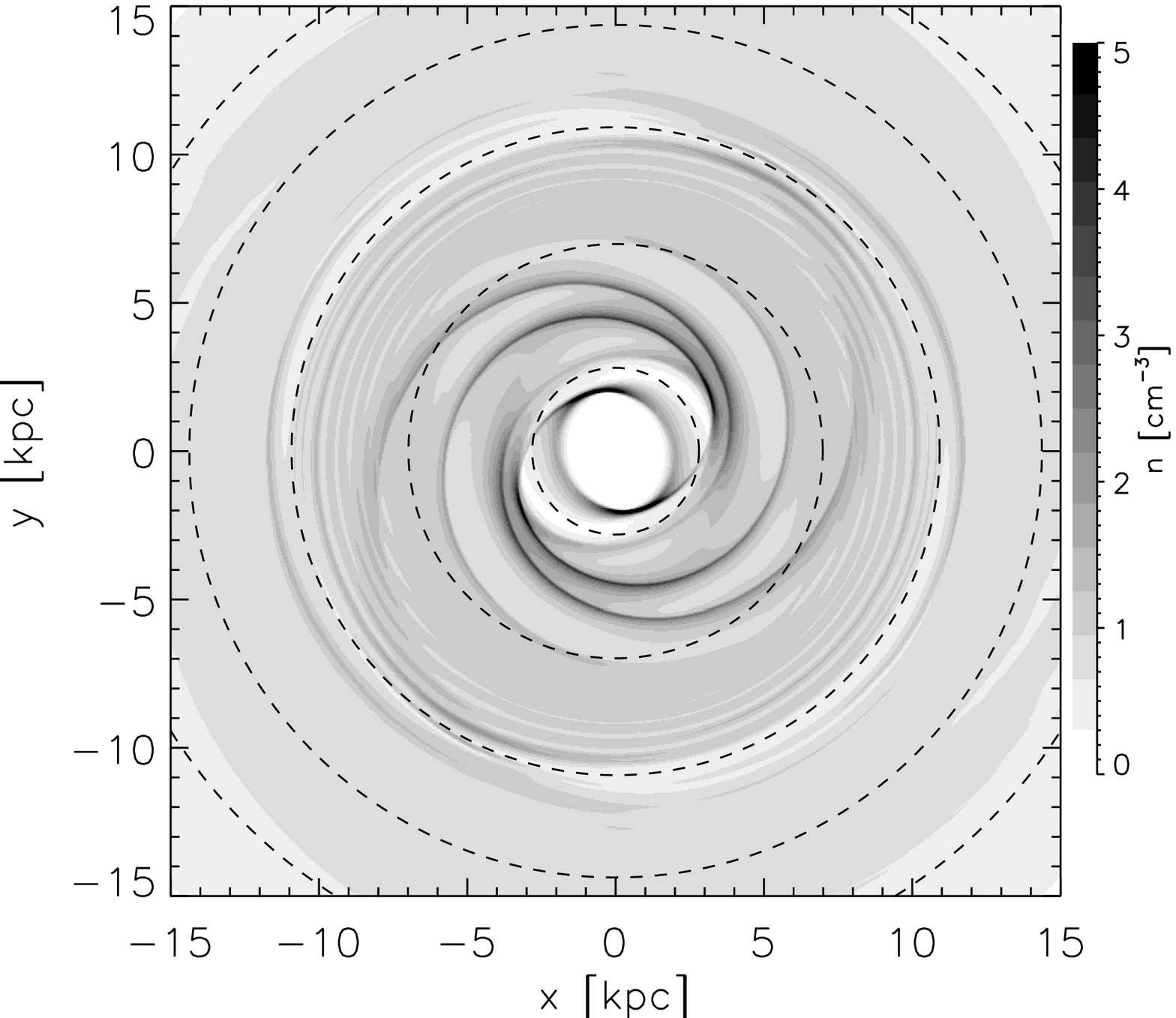}\\
  \includegraphics[width=0.85\hsize]{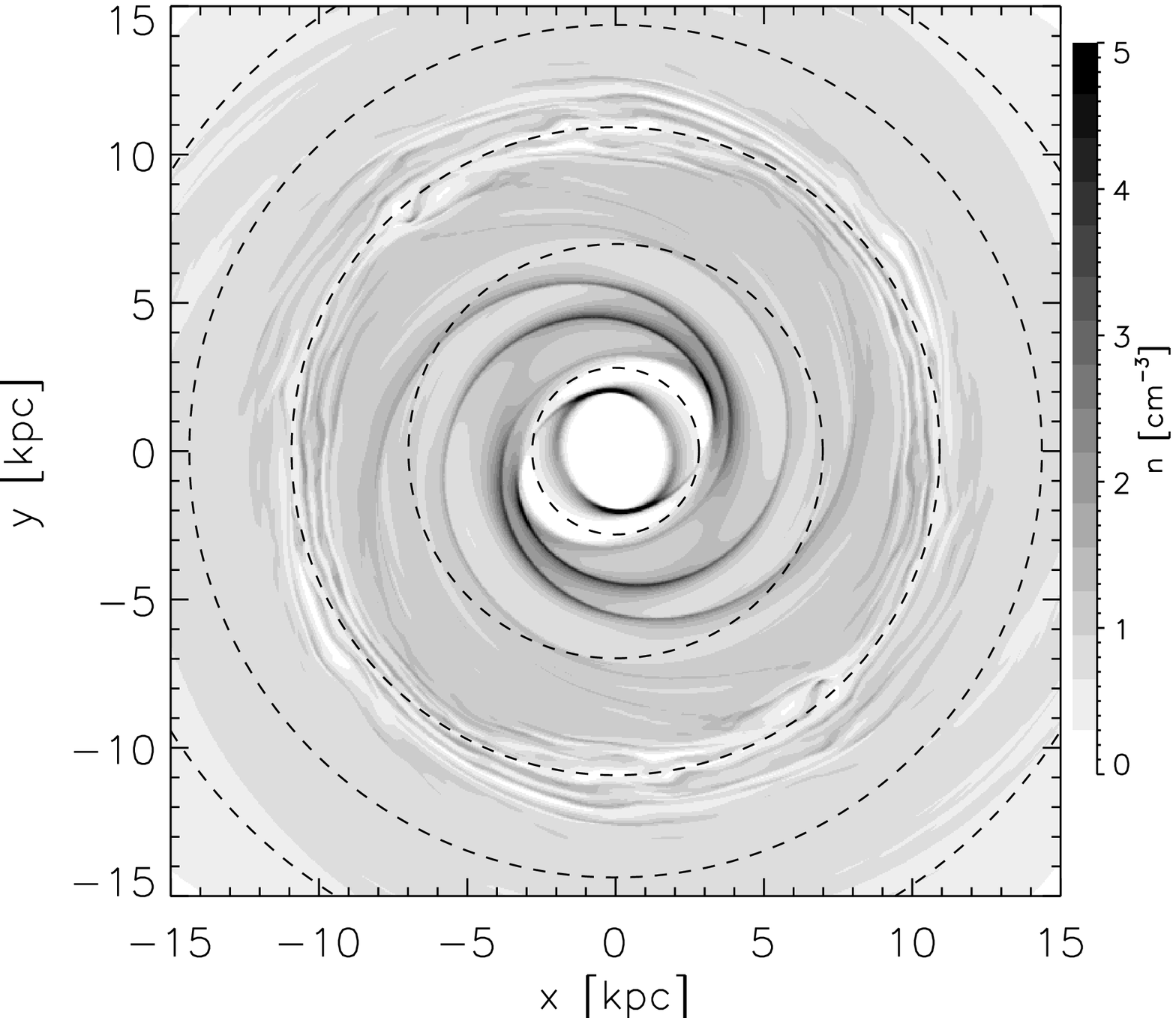}
  \caption{Density distribution after
  $1\Gyr$ ({\it top}), $2.5\Gyr$ ({\it middle}), and $4\Gyr$ ({\it
  bottom}).
  Although the computational grid extends to $r=22\kpc$,
  only the inner region is shown.
  The dashed circles show the position of the inner Lindblad, 4:1,
  corotation, -4:1 and outer Lindblad resonances.
  Gas rotates in the clockwise direction (counterclockwise
  outside the corotation resonance).
  }
  \label{fig:simulation}
\end{figure}

\section{Numerical model}
\label{sec:model}

For the purpose of this work we have produced MHD simulations under
the effect of a detailed model of the background galactic potential.
We describe next the potential model and the performed simulations.

\begin{table*}
  \centering
  \begin{minipage}{0.6\textwidth}
    \caption{Non-axisymmetric galactic model \citep{pic03}}
    \begin{tabular}{@{}lclcc@{}}
      \hline
      Parameter                                 & Value       & Reference\\
      \hline
      Spiral arms locus               & Bi-symmetric (Logthm) & \citet{chu09}\\
      Spiral arms pitch angle                   & $15.5\degr$ & \citet{dri00} \\
      Spiral arms external limit                & $12\kpc$    & \citet{dri00} \\
      Spiral arms: exp. with scale-length       & $2.5\kpc$   & Disk based \\
      Spiral arms force contrast                & $\sim10\%$  & \citet{pat91} \\
      Spiral arms pattern speed ($\Omega_{P}$)  & $20\kms\kpc$& \citet{mar04}\\
      \hline
    \end{tabular}
    \label{tab.param}
  \end{minipage}
\end{table*}

\subsection{The Galactic Background Model}

For the orbital study we have employed a detailed galactic
semi-analytic model, resembling a Milky-Way-like
potential.
Rather than
using a simple ad hoc model for a three-dimensional spiral
perturbation (such as a cosine function), we constructed a
three-dimensional mass distribution for the spiral arms and
derive their gravitational potential and force fields from previously
known results in potential theory. The spiral arms distribution
consists of inhomogeneous oblate spheroids superposed along a given
spiral locus.
It is physically simple, with continuous derivatives and density
laws. In principle, the dimensions and mass density of the oblate spheroids
will depend on the type of spiral arms that are modeled, gaseous or stellar.
The model, called {\tt PERLAS}, is described in detail in
\citet{pic03}.
Comparisons of the model with other models have been
already published, in particular for the most relevant parts of our
study, the spiral arms perturbation \citep{ant09,mar04,pic03}.
The corresponding observationally motivated
parameters of the model used for this work are presented in Table
\ref{tab.param}. The spiral perturbation is a 3D steady two-armed
model that traces the locus reported by \citet{dri01}, using K-band
observations. The solar radius, at $8.5\kpc$, is close to the spiral
arms 4:1 resonance. The mean force ratio between
the arms and the axisymmetric background is around 10\%, which
is in agreement
with the estimations by \citet{pat91} for Milky Way type
galaxies.
At the solar circle, the radial component of the
force due to the spiral arms is $4.4\%$ of the mean axisymmetric
background. The self-consistency of the spiral arms has been tested
through the reinforcement of the spiral potential by the 
stable periodic orbits
\citep{pat91,pic03}.

\subsection {The MHD Setup}

The initial setup of the MHD simulations consists of a gaseous disk
with a density profile given by
$n(r) = n_0 \exp[-(r-r_0)/r_d]$, where
$n_0=1.1\pcc$, $r_0=8\kpc$, and $r_d=15\kpc$.
The gas behaves isothermally with a temperature $T=8000\degK$ and
is permeated by a magnetic field, initially in the azimuthal
direction, with intensity $B(r) = B_0 \exp[-(r-r_0)/r_B]$
where $B_0 = 5\muG$ and $r_B=25\kpc$.
The disk is initially in rotational equilibrium between the
centrifugal force, the thermal and magnetic pressures, magnetic
tension and a background axisymmetric galactic potential \citep{all91}
consisting of a bulge, stellar disk and halo.

This equilibrium is perturbed by the {\tt PERLAS} spiral arm
potential, which rotates with a pattern speed of
$\Omega_P=20\kms\kpc^{-1}$.
This pattern speed and background potential place the inner
Lindblad, 4:1, corotation, -4:1, and outer Lindblad resonances
at $2.81,\, 6.98,\, 10.92,\, 14.37$ and $17.72\kpc$, respectively.

We solved the MHD equations with the {\tt ZEUS} code
\citep{sto92a,sto92b}, which is a finite difference, time explicit,
operator split, Eulerian code for ideal magnetohydrodynamics.
We used a 2D grid in cylindrical geometry, with $R\in[1.5\kpc,22\kpc]$
and a full circle in the azimuthal coordinate, $\phi$, using
$750\times1500$ grid points.
Both boundary conditions in the radial direction were outflowing.
All the calculations are performed in the spiral pattern reference
frame.
No self-gravity of the gas was considered.

The simulation starts with the gas in circular orbits in equilibrium
with the background galactic potential, thermal and magnetic pressure
gradients and magnetic tension.
After the perturbation is activated, the gas very rapidly settles
into a spiral pattern, with two arms
(plus an oval-shaped ring\footnote{This ring at $r < 2\kpc$
falls in a region where the dynamics should be dominated by a
galactic bar, which is not included in this work. Therefore, it is
disregarded.})
inside the inner Lindblad
resonance (ILR) and four arms outside, up to the 4:1
resonance (see fig.~\ref{fig:simulation}).
The inner pair of arms follows a $\sim9\degr$ pitch angle, while the
two pairs outside the ILR follow a $\sim9\degr$ and $\sim13\degr$ pitch
angles;
in contrast, the perturbing arm potential follows a $15.5\degr$ pitch.
The exact position and pitch angle of the gaseous spiral arms oscillate
slightly around the above quoted values.

It is noticeable the presence of an MHD instability at corotation
radius \citep[further described in][]{mar13},
which begins to develop $2.8\Gyr$ after the start of the simulation.
This instability is present in a variety of simulations with
different magnetic field intensities, and absent in purely
hydrodynamical ones.
With exception of this instability,
absence of a magnetic field does not affect our conclusions.

\begin{figure}
  \includegraphics[width=0.90\hsize]{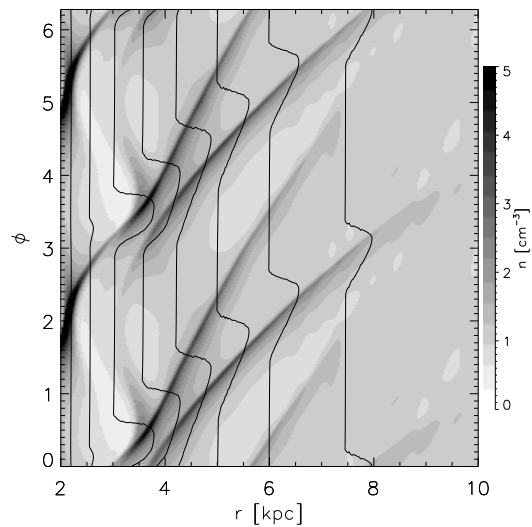}
  \caption{
    Comparison of the stellar (contours) and gaseous densities
    (grayscale) in the simulation.
    The stellar density is that corresponding to the background
    axisymmetric galactic disk plus the (imposed) spiral arms
    potential, while the gas density corresponds to
    to that of a snapshot taken $1\Gyr$ after the start of
    the simulation.
    Since the model corresponds to a trailing spiral, the gas flows
    down from the top of the plot.
    The differences in number and position of the stellar and
    gaseous arms are apparent.
  }
  \label{fig:densidades}
\end{figure}

Figure~\ref{fig:densidades}
compares the imposed stellar density with the response gas density at
a time $1\Gyr$ after the start of the simulation.
The stellar density is obtained by taking the Laplacian of the
potentials corresponding to the
axisymmetric background disk and spiral arm distribution, as given
by the {\tt PERLAS} model.
In this figure, the differences in position and pitch of stellar and
gaseous arms are apparent.

\begin{figure}
  \includegraphics[width=0.90\hsize]{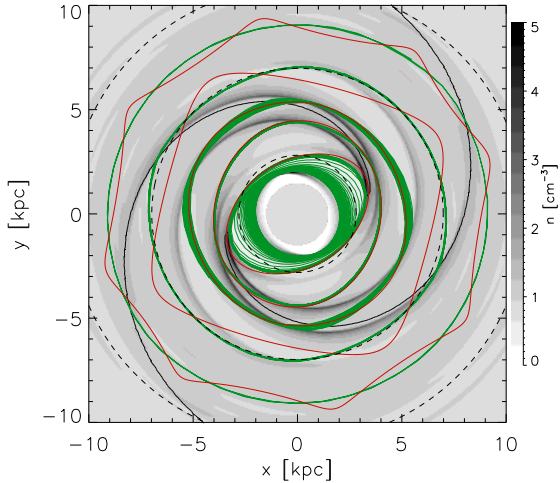}
  \caption{
    Comparison of some 
    stellar ({\it red}) and gaseous ({\it green}) orbits near major
    resonances.
    The location of the inner Lindblad (ILR), 4:1 and corotation are
    indicated by dashed circles.
    The gaseous orbit near the ILR starts similar to the stellar
    orbit, but it rapidly decays to a ring.
    Around other resonances, the gas follows a near circular orbit,
    even as the stellar orbits experience large radial excursions.
    As reference, the gas density ({\it grayscale}) and the
    locus of the stellar arms ({\it black line}) are also shown.
  }
  \label{fig:junta_orbitas}
\end{figure}


\section{Orbits in the spiral potential}
\label{sec:orbits}

Figure~\ref{fig:junta_orbitas} shows a comparison of stable periodic
stellar and gaseous orbits in the background axisymmetric+spiral
galactic potential.
That figure also shows the positions of the major resonances and the
locus of the spiral arms as defined in the {\tt PERLAS} model.
Since {\tt ZEUS} is an Eulerian code, it does not provide the actual
path a gas parcel follows.
Therefore, in order to reconstruct a gaseous orbit, we
integrated the velocity field interpolating in space and time between
data files for an array of initial positions, following the
trajectories of individual gas parcels in the resulting velocity
field.
The integration was performed over $4\Gyr$, starting $1\Gyr$ into the
simulation in order to avoid the initial settling of the gas into
the spiral pattern.
If during that time the gas parcel reaches either of the numerical
boundaries (at $R=1.5$ and $22\kpc$), we stop the integration.
The stellar orbits were obtained by direct integration in the
potential, locating the periodic ones using a Newton-Raphson
algorithm.

When comparing the stellar and gaseous orbits presented in
Figure~\ref{fig:junta_orbitas}, it is apparent that some of them are
similar, but also some may be quite different, even when
both gas and stars are subjected to the same gravitational potential.

\begin{figure*}
  \includegraphics[width=0.24\hsize]{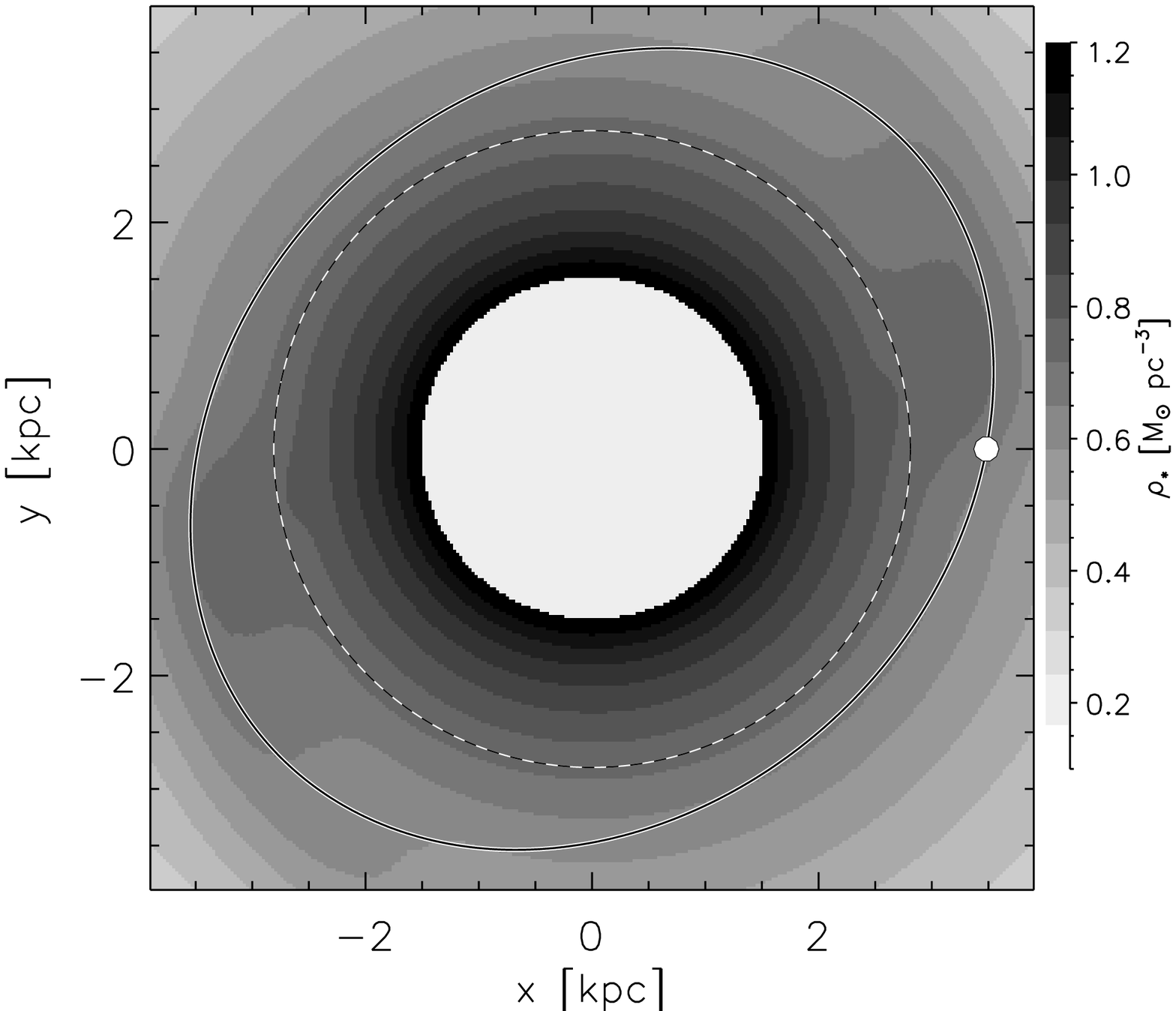}
  \includegraphics[width=0.24\hsize]{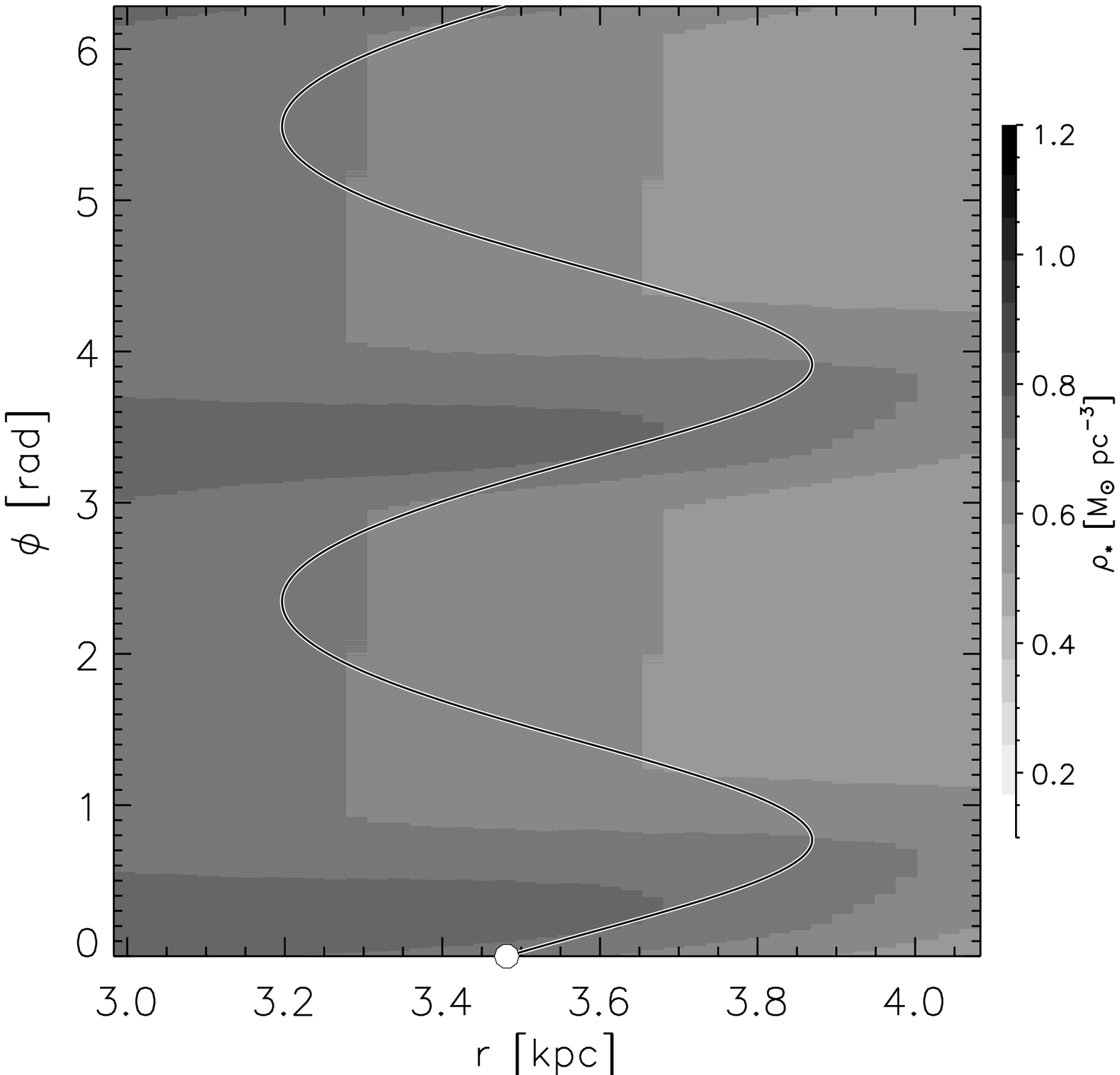}
  \includegraphics[width=0.24\hsize]{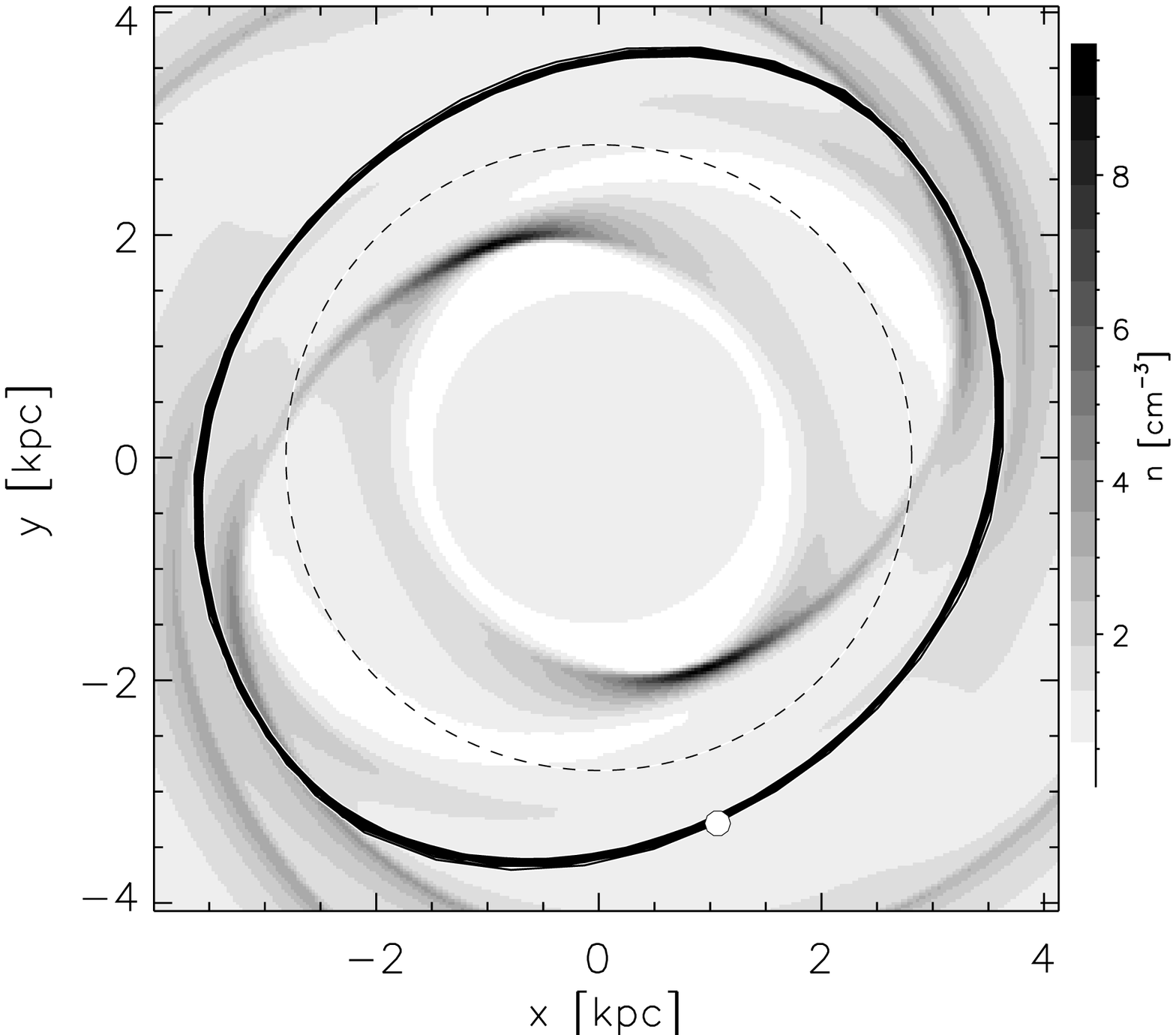}
  \includegraphics[width=0.24\hsize]{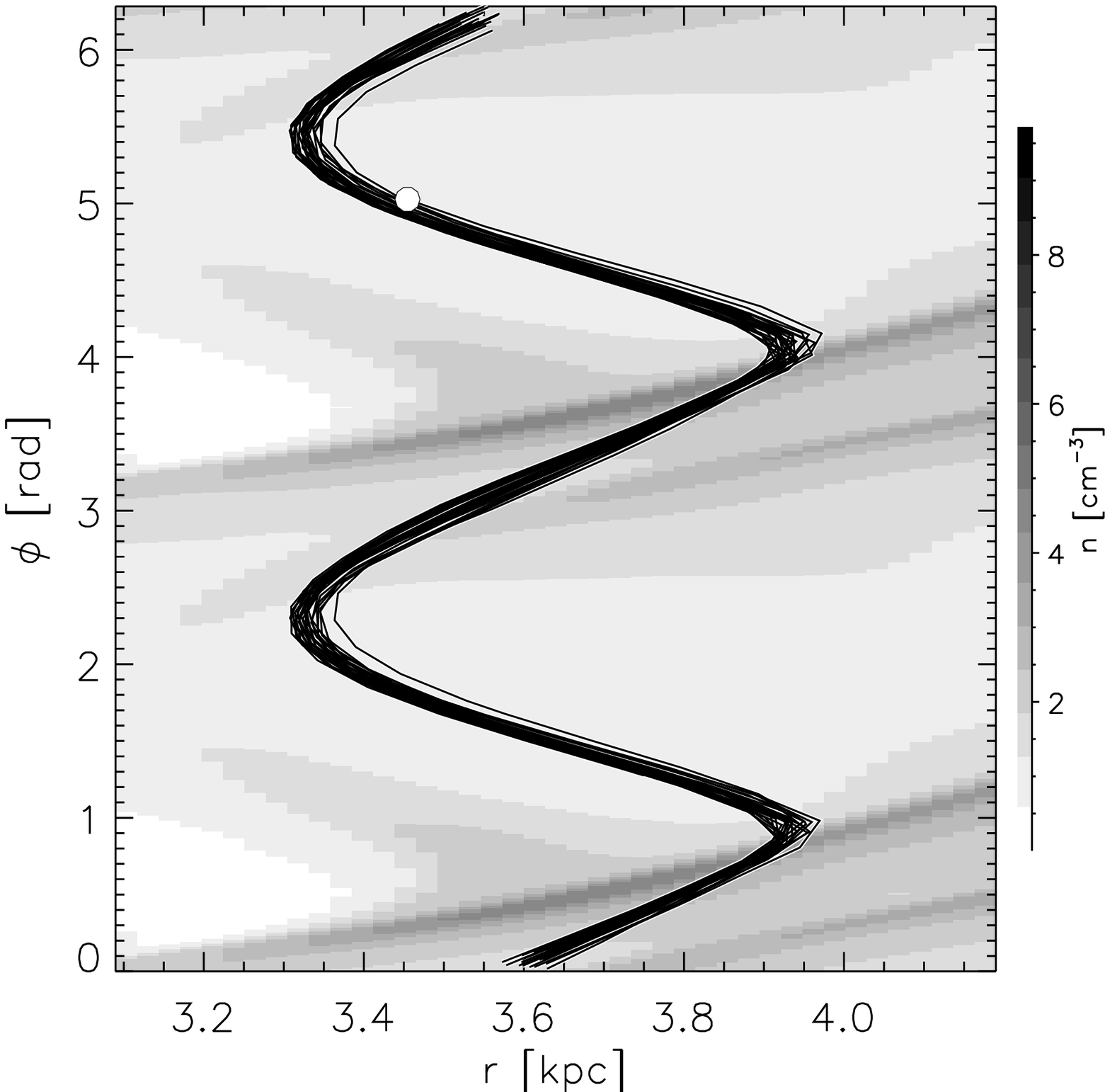}
  \\
  \includegraphics[width=0.24\hsize]{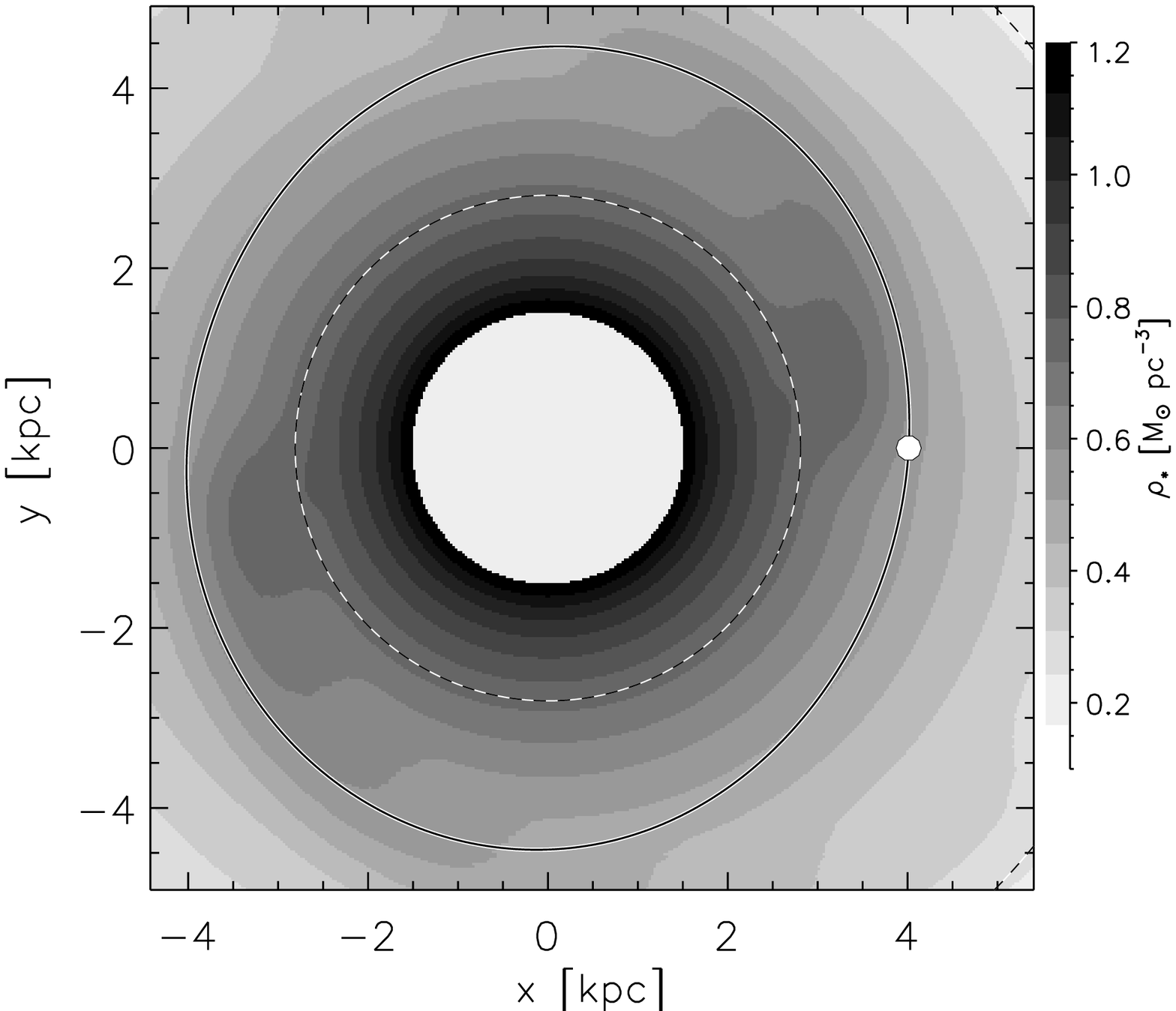}
  \includegraphics[width=0.24\hsize]{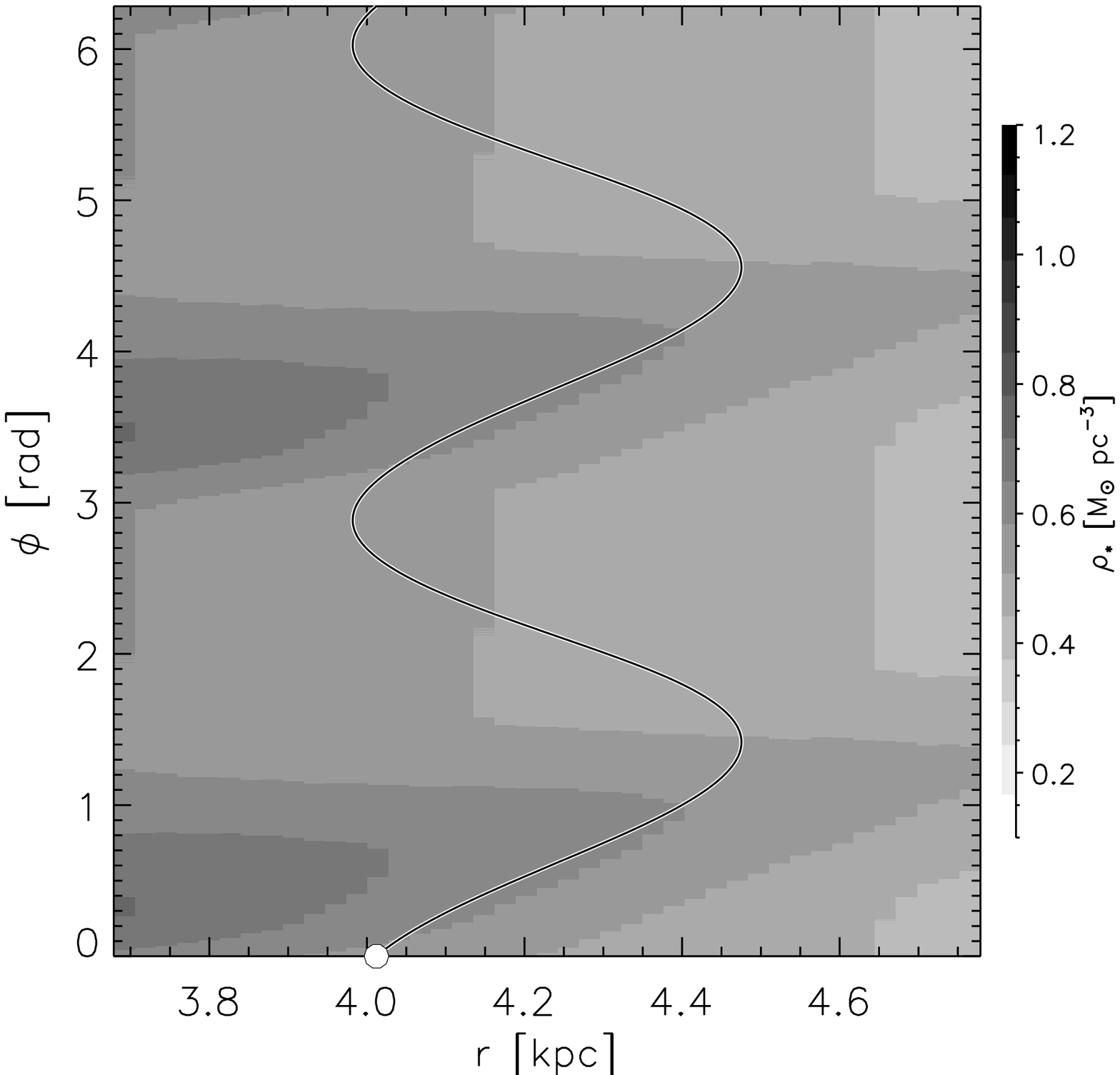}
  \includegraphics[width=0.24\hsize]{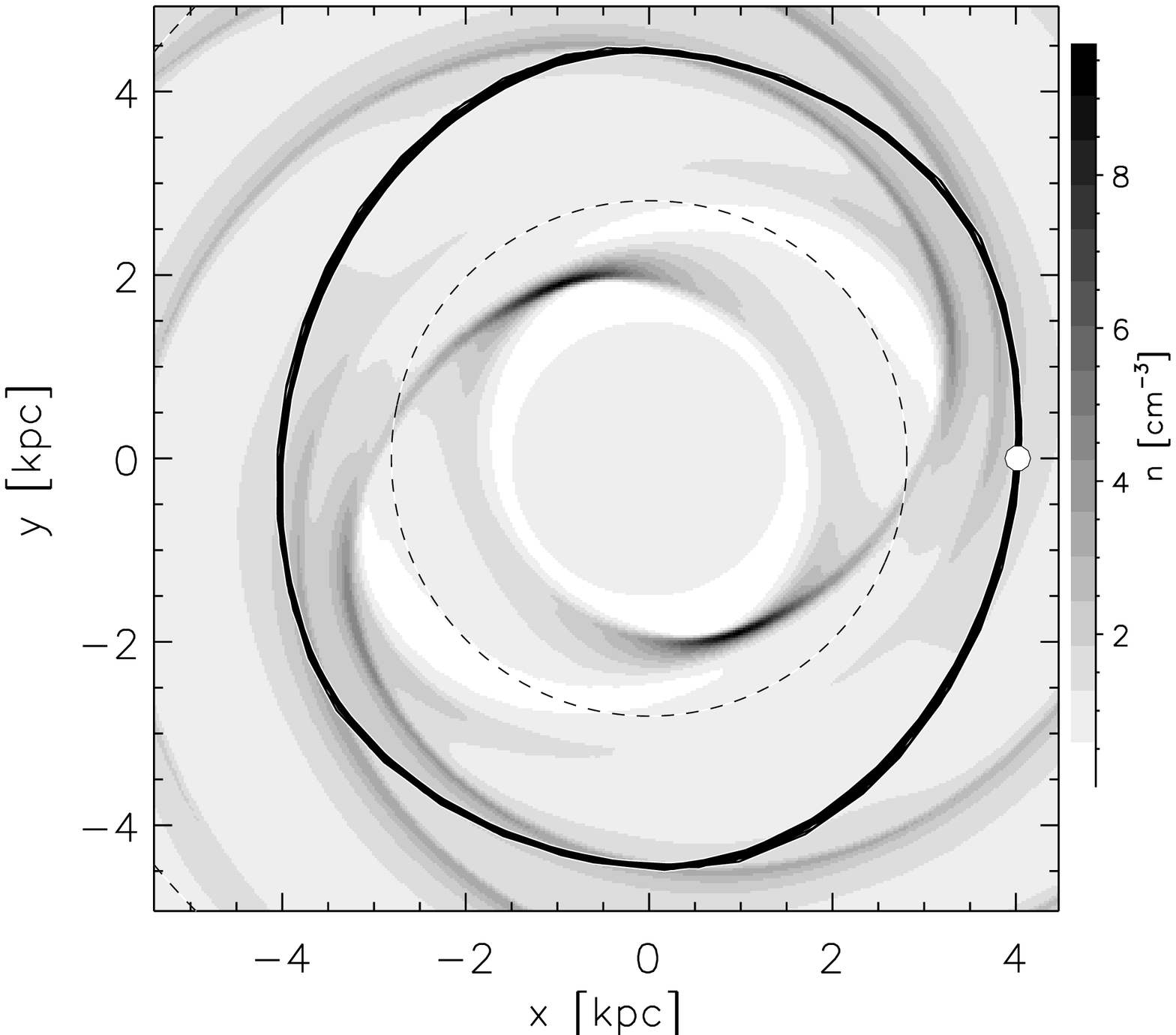}
  \includegraphics[width=0.24\hsize]{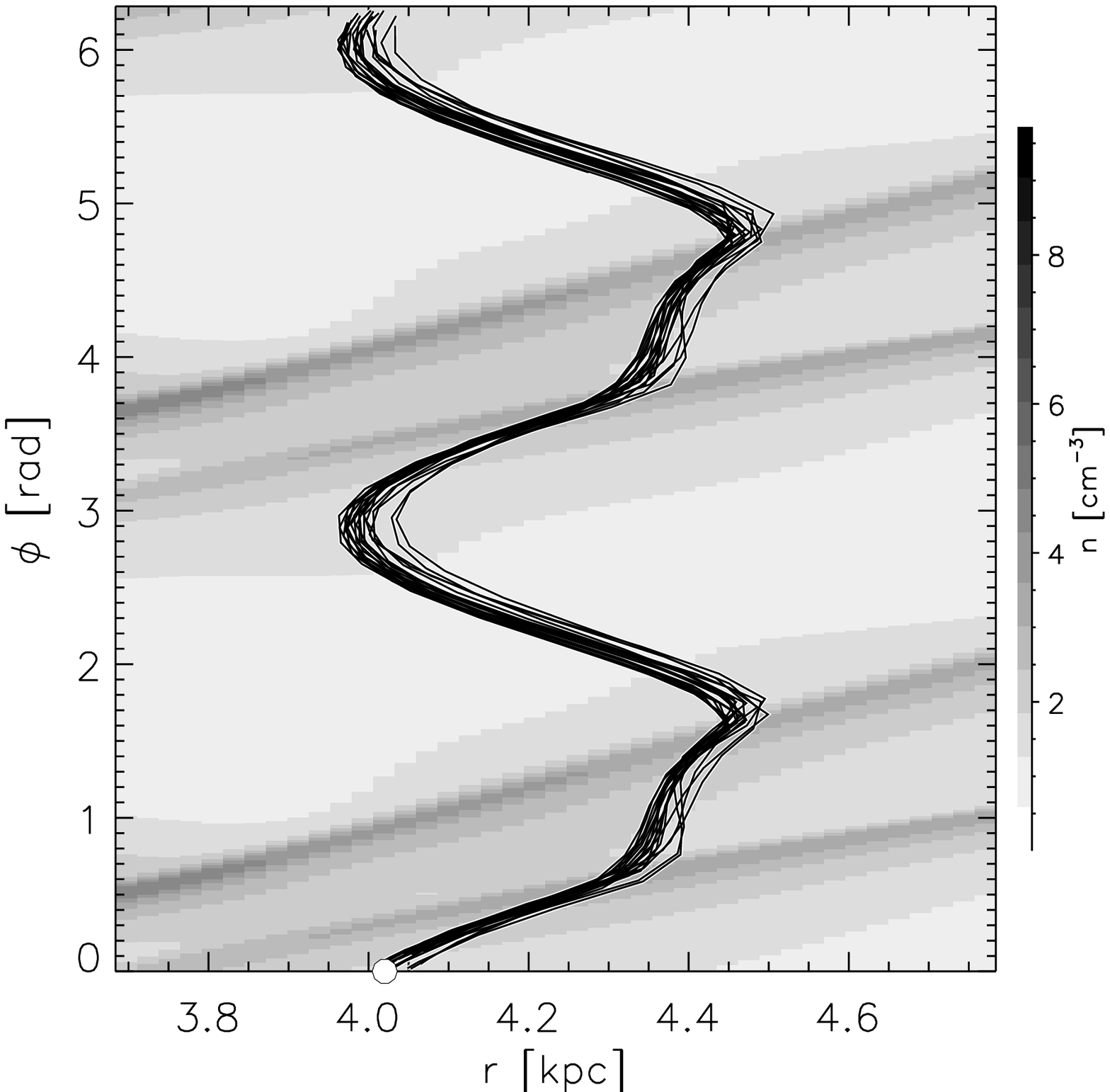}
  \\
  \includegraphics[width=0.24\hsize]{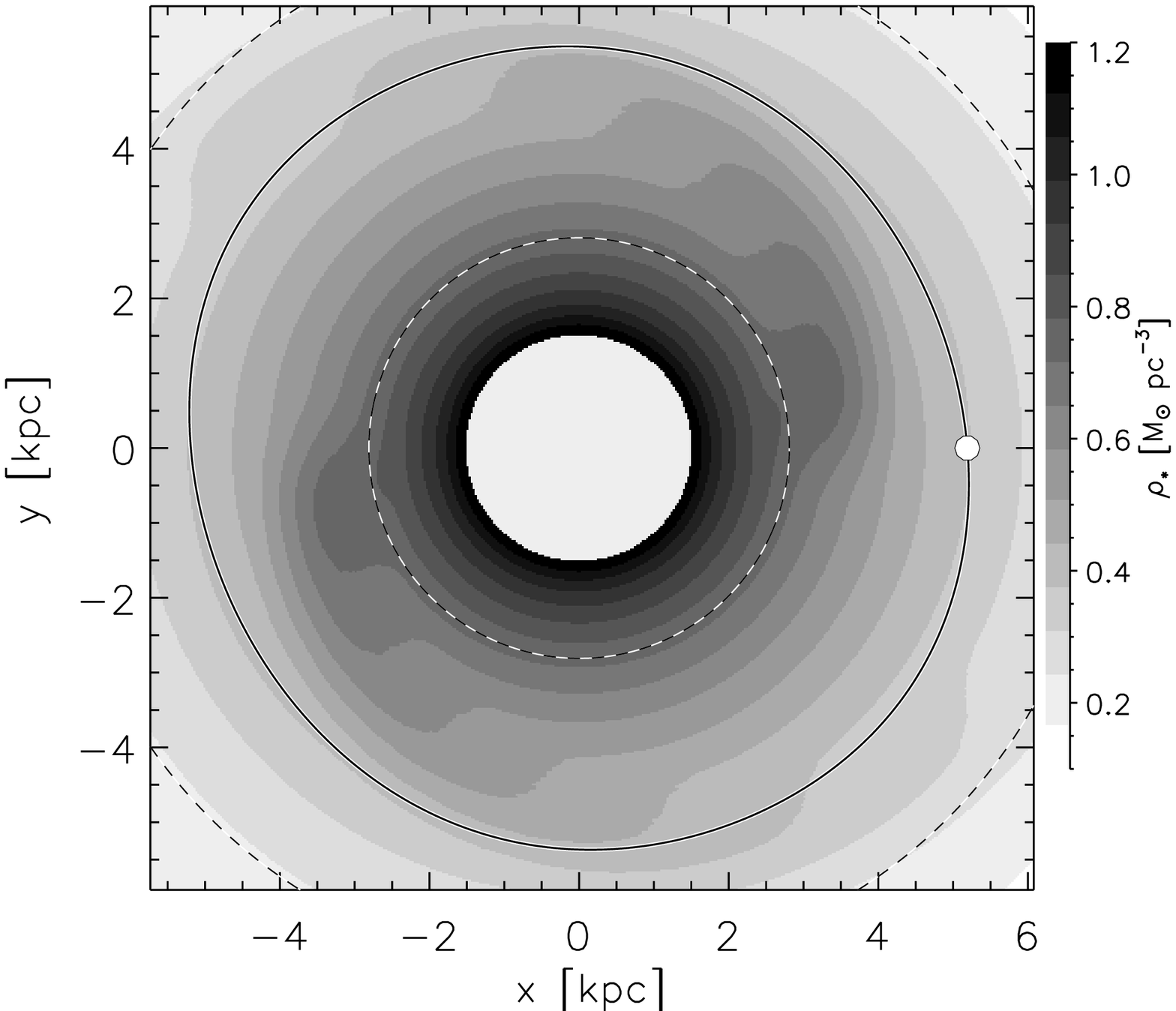}
  \includegraphics[width=0.24\hsize]{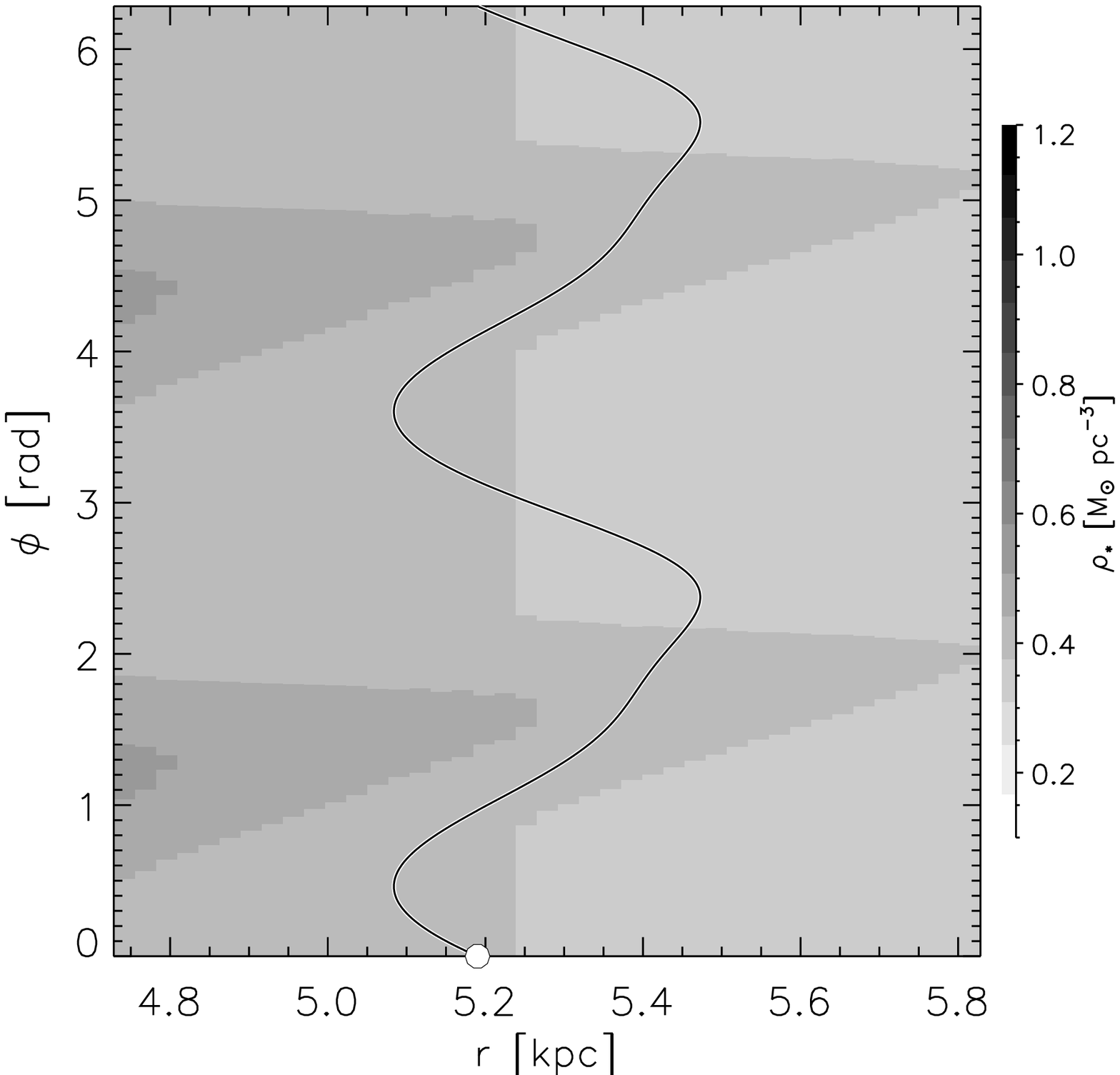}
  \includegraphics[width=0.24\hsize]{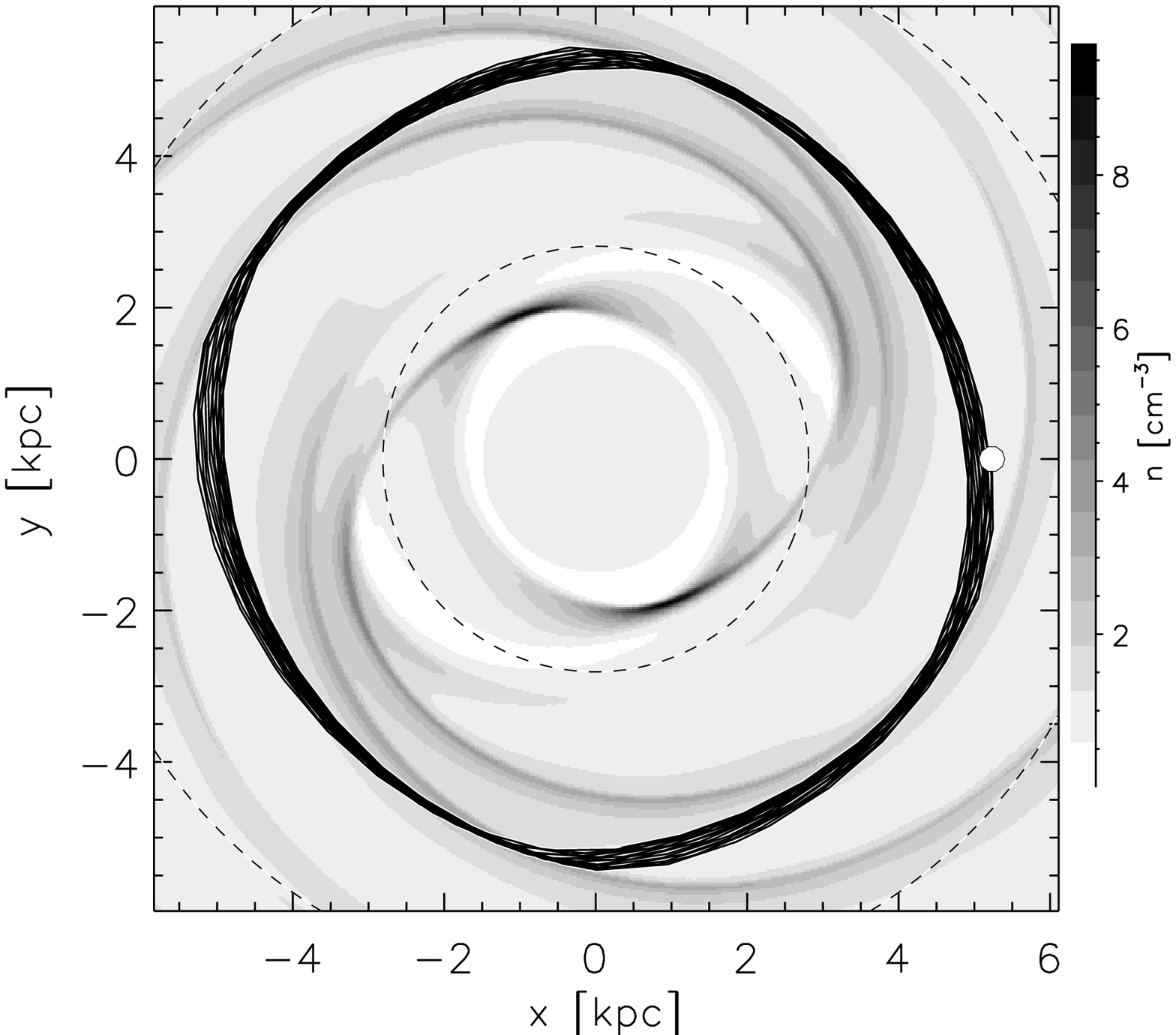}
  \includegraphics[width=0.24\hsize]{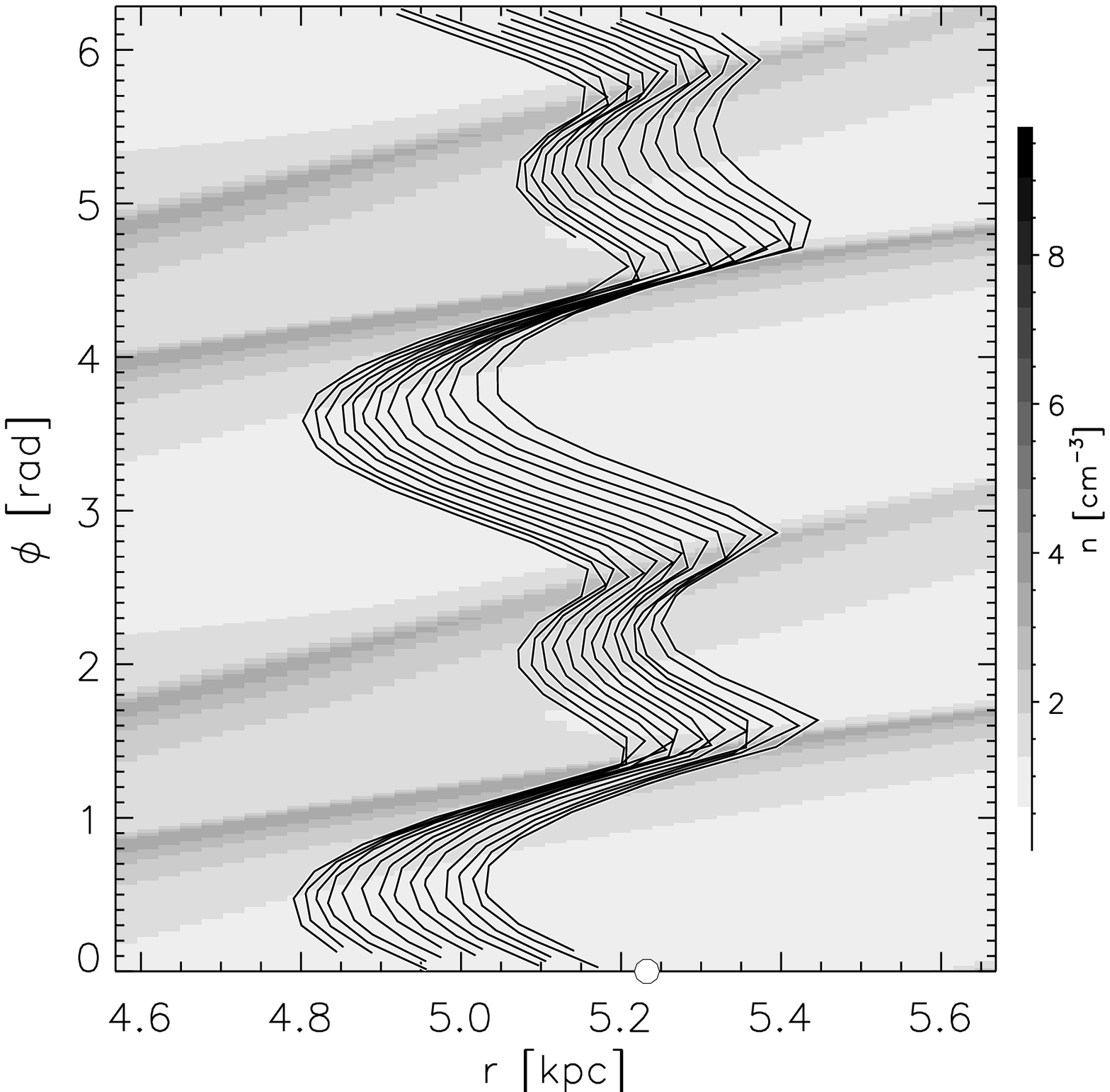}
  \\
  \includegraphics[width=0.24\hsize]{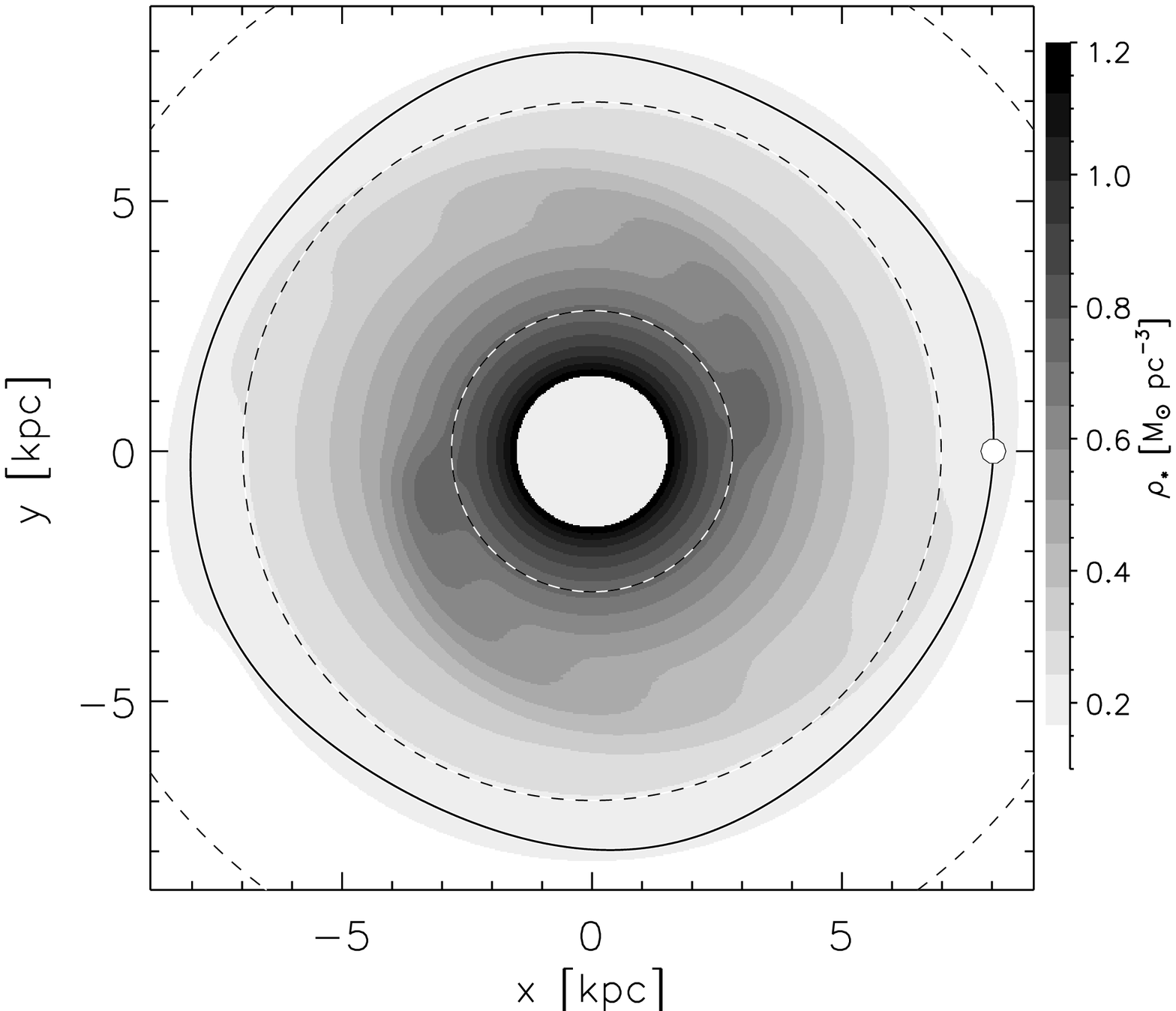}
  \includegraphics[width=0.24\hsize]{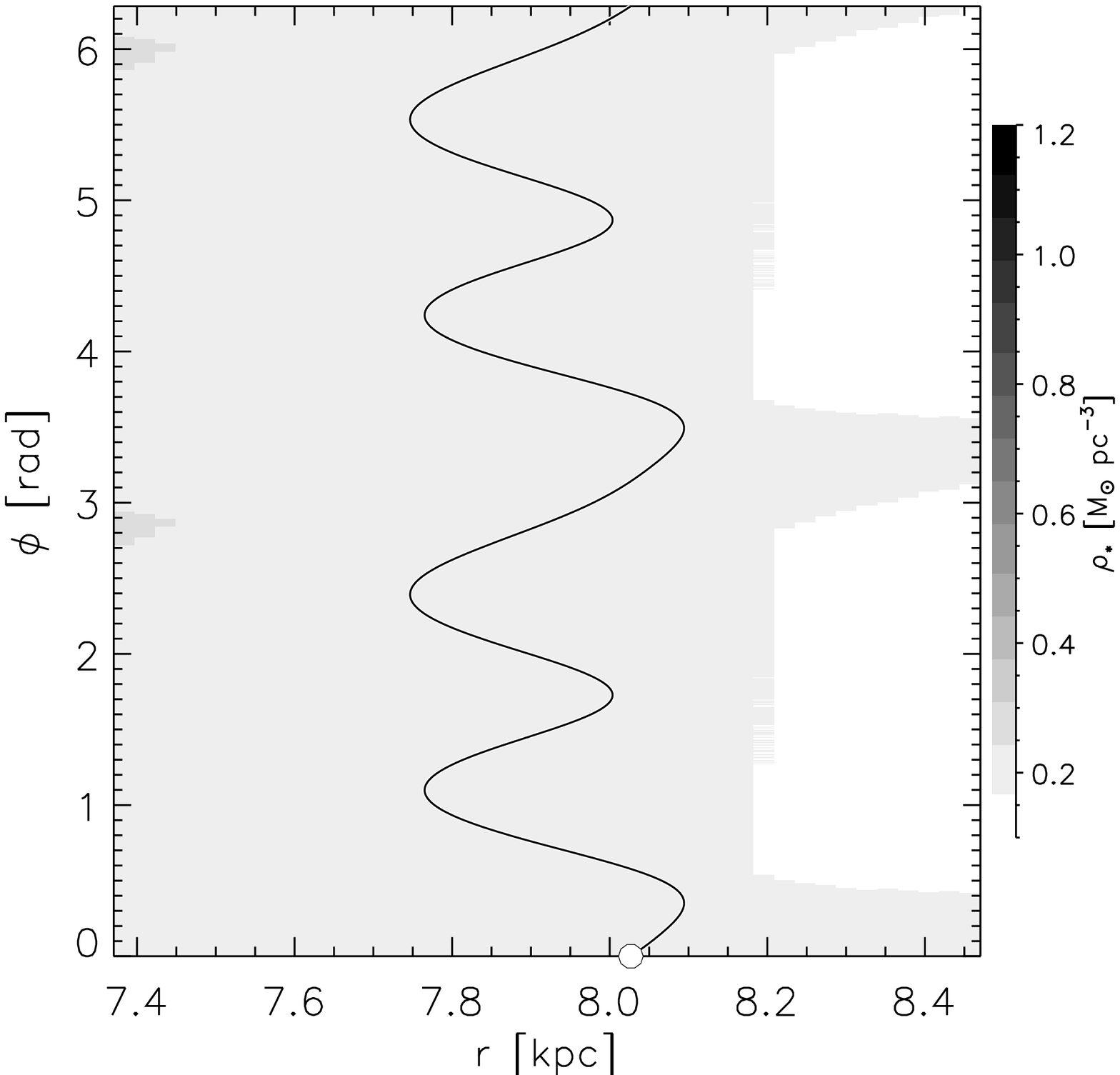}
  \includegraphics[width=0.24\hsize]{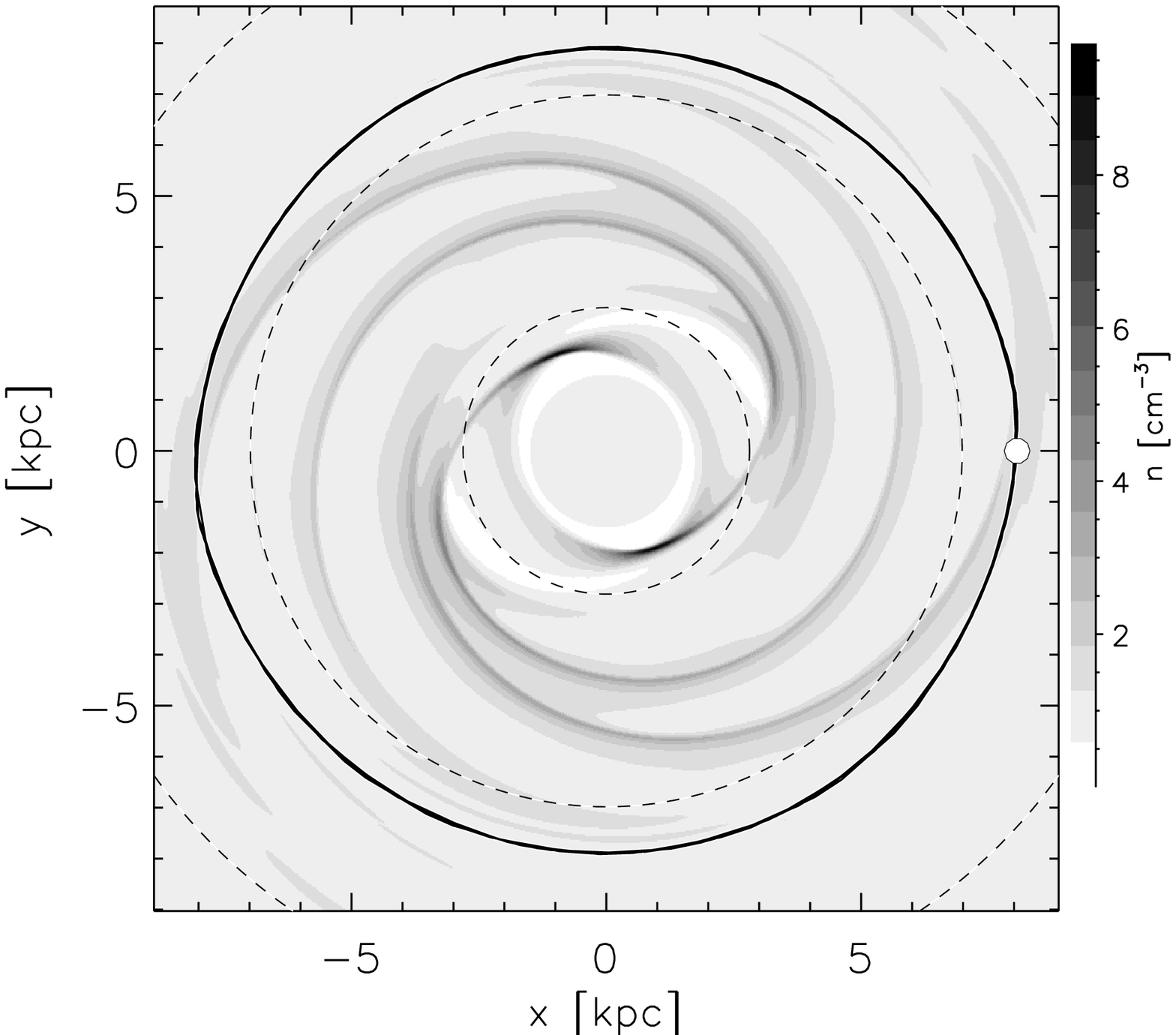}
  \includegraphics[width=0.24\hsize]{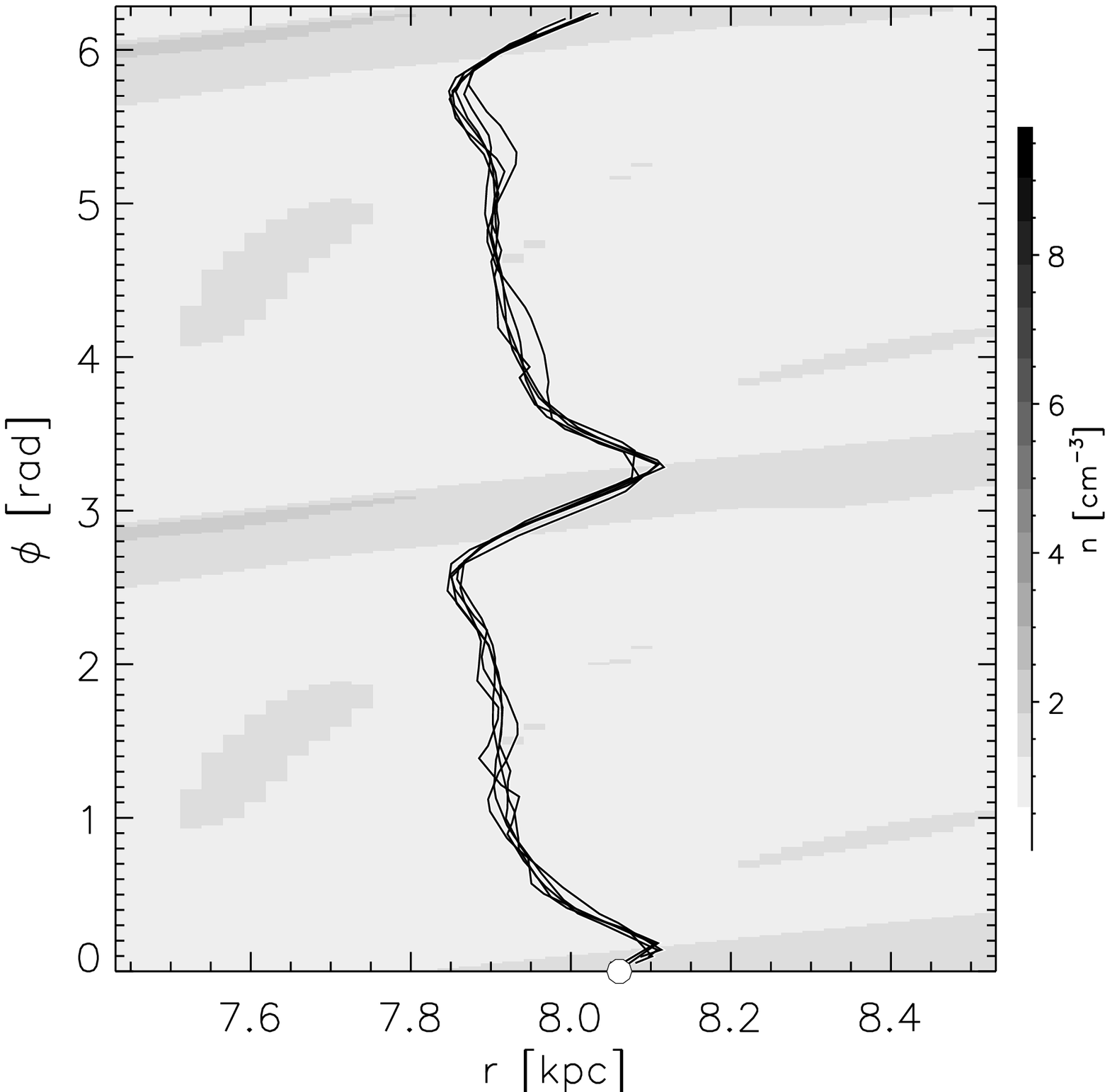}
  \caption{
    Comparison of stellar and gaseous orbits between resonances.
    The first two columns show stellar orbits in cartesian ({\it
    first column}) and polar coordinates ({\it second column}) plotted
    over the stellar mass density ({\it grayscale}) corresponding to
    the {\tt PERLAS} potential model.
    The last two columns show gaseous orbits, also in cartesian and
    polar coordinates, plotted over the gas density ({\it
    grayscale}) averaged over the $4\Gyr$ period of the orbit
    integration.
    In all cases, the initial position of the integration is marked
    by the small open circle.
    Since the model corresponds to a trailing spiral, the sense of
    the gaseous orbits is clockwise in the cartesian coordinate
    plots, and down from the top in the polar coordinate ones;
    only stellar orbits with the same sense of rotation were
    considered.
    The position of the inner Lindblad, 4:1 and 
    corotation resonances is shown ({\it dashed circles}).
  }
  \label{fig:orbits_far_res}
\end{figure*}

Stars and gas follow orbits that are similar
in the regions between the resonances.
In Figure~\ref{fig:orbits_far_res}, we show some of the orbits
calculated in similar regions; we also show the stellar mass and gas
densities for comparison with the orbit.
In the case of the stellar orbits,
it is noticeable how the familiar oval shape
(associated with the ILR) gradually morphs into a rounded square
as the orbit approaches the 4:1 resonance.
These last orbits are less supportive of the spiral pattern, in
accordance with analytical theory.

\begin{figure*}
  \includegraphics[width=0.24\hsize]{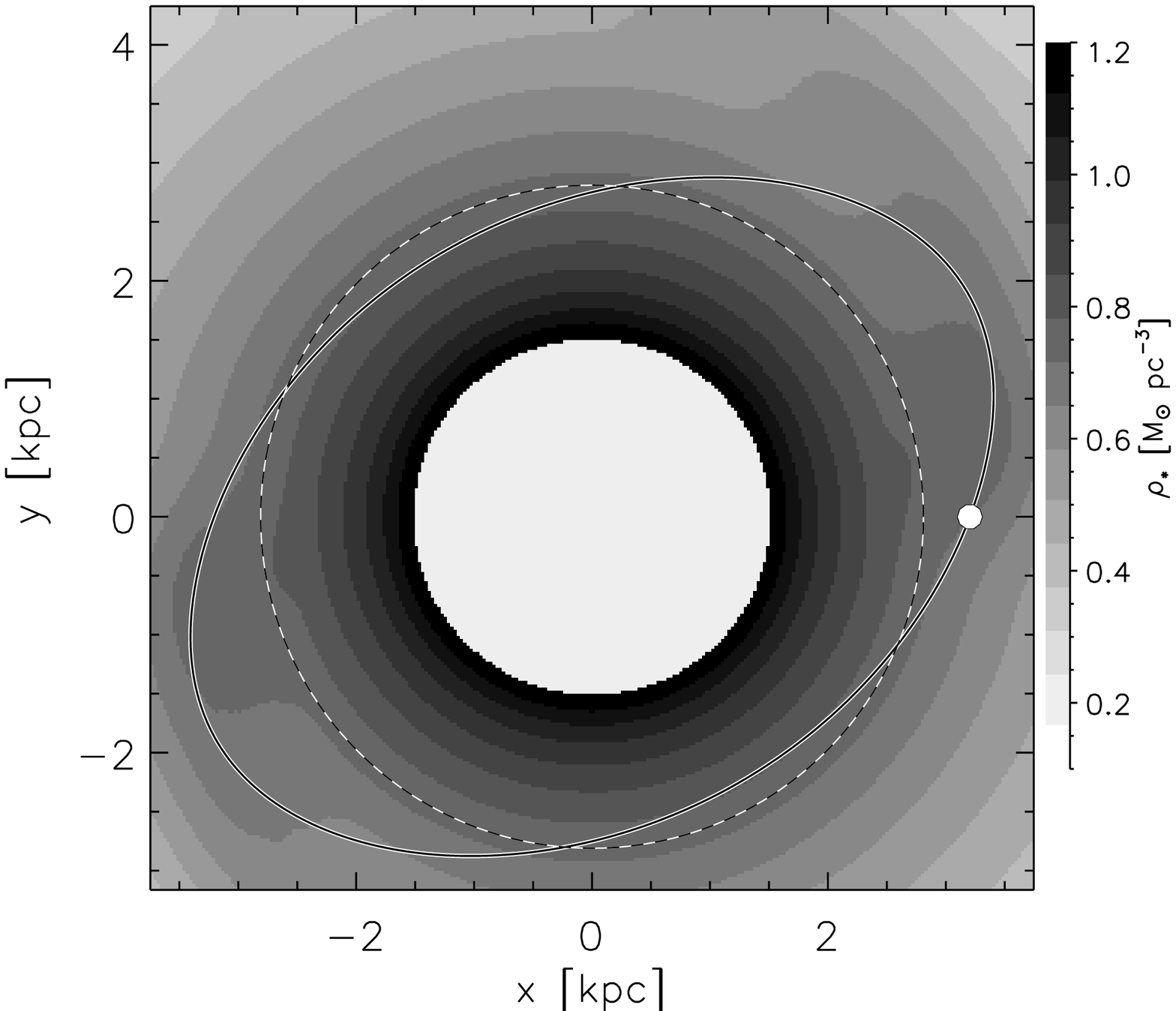}
  \includegraphics[width=0.24\hsize]{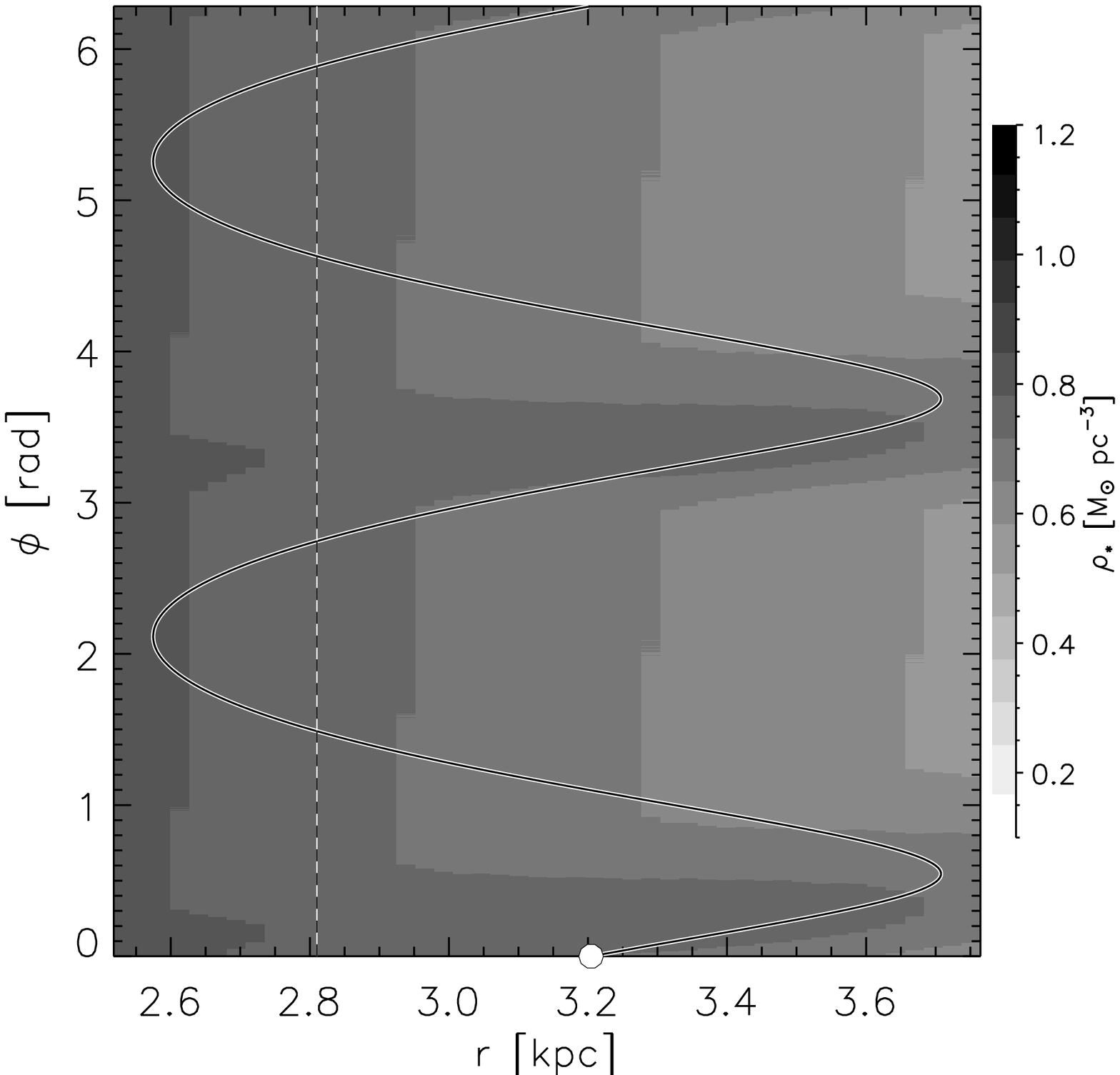}
  \includegraphics[width=0.24\hsize]{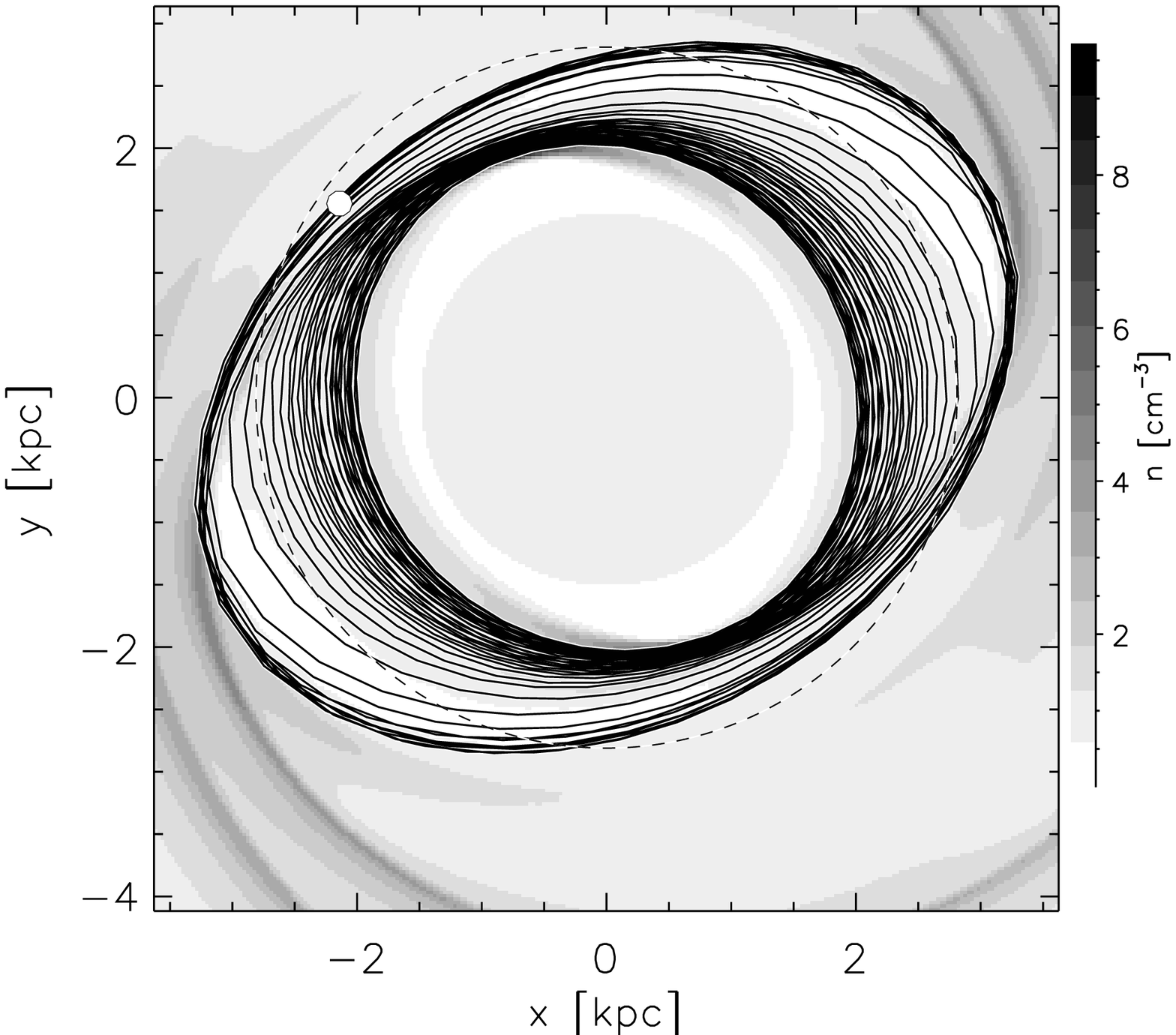}
  \includegraphics[width=0.24\hsize]{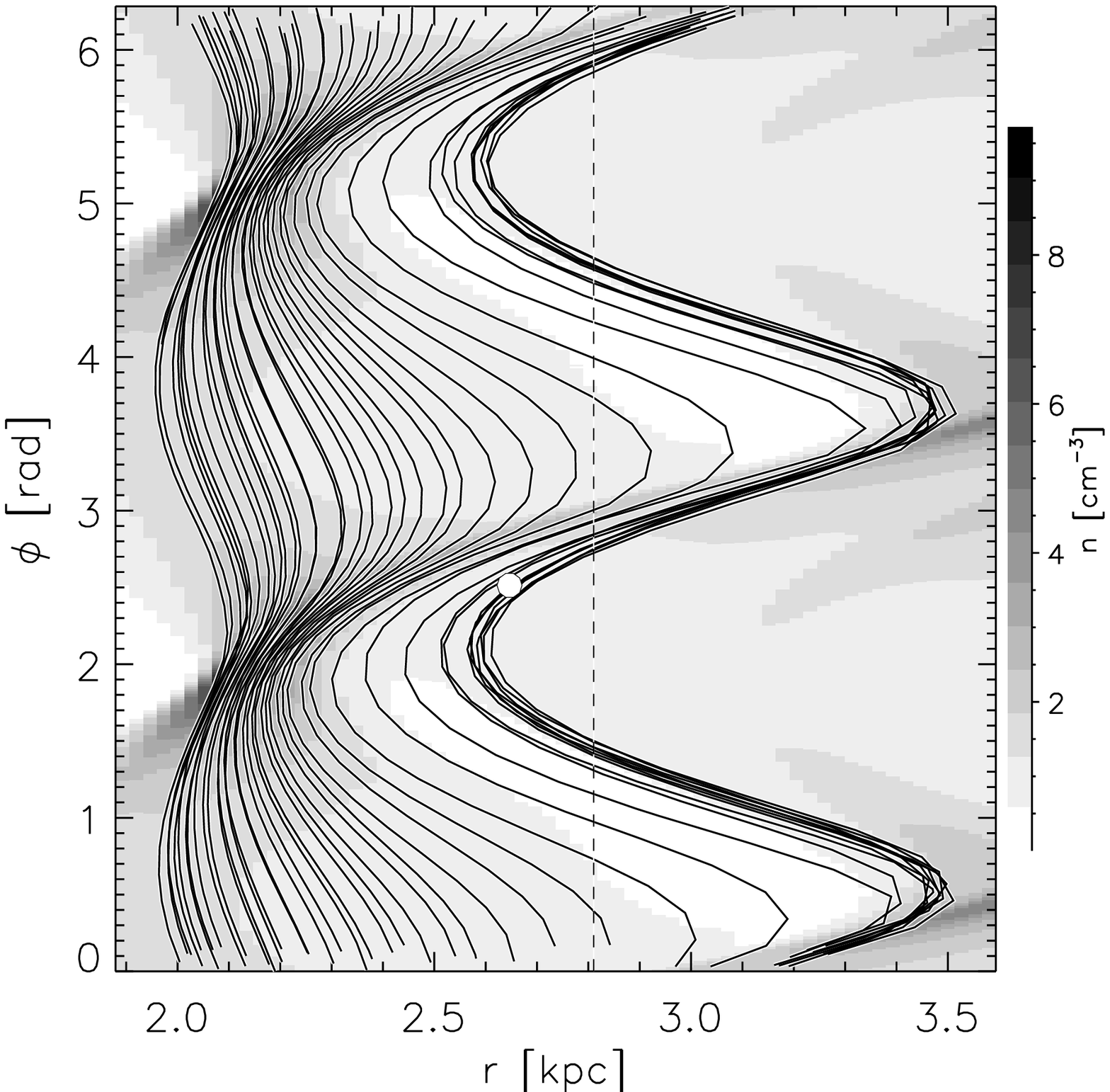}
  \\
  \includegraphics[width=0.24\hsize]{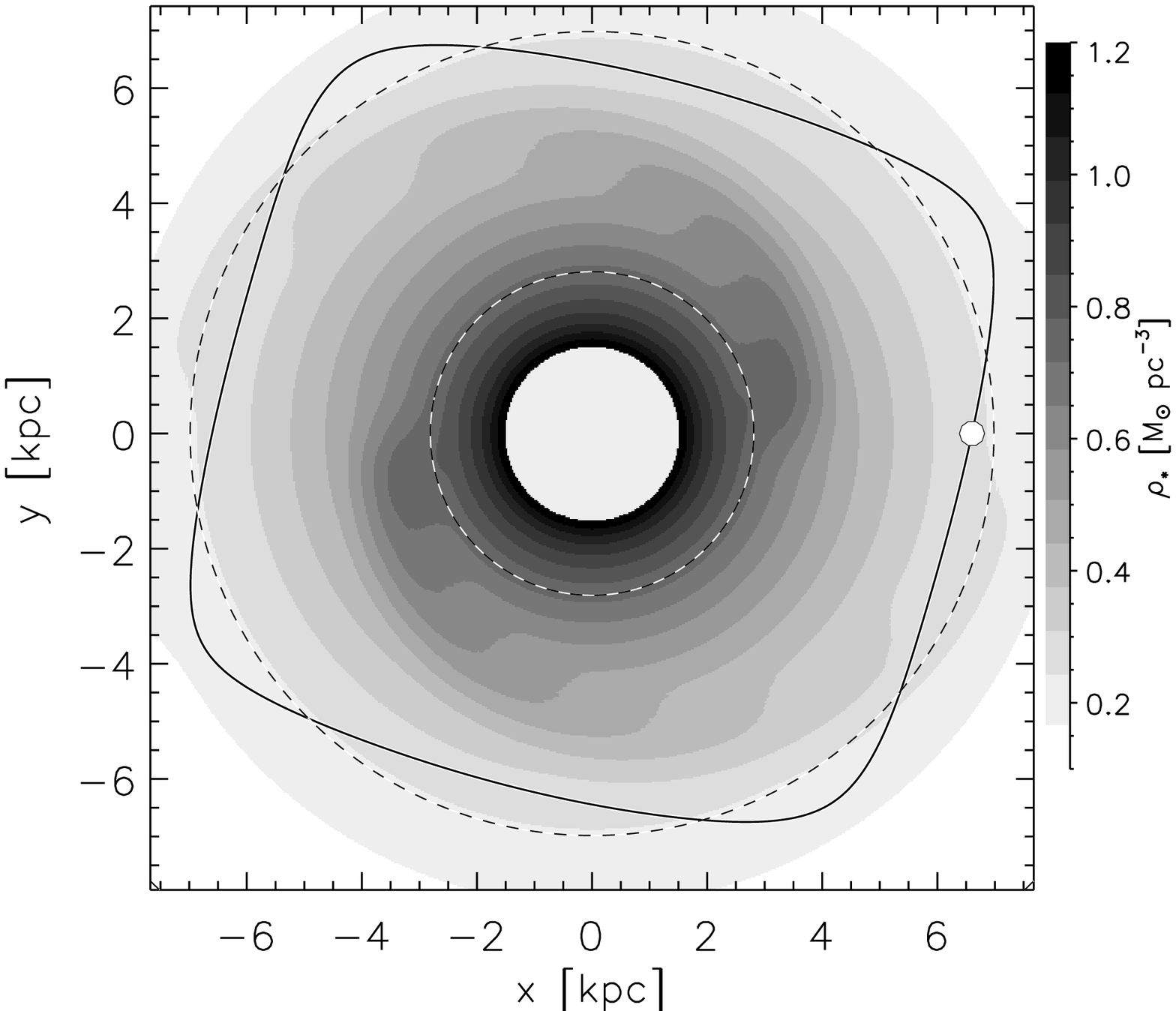}
  \includegraphics[width=0.24\hsize]{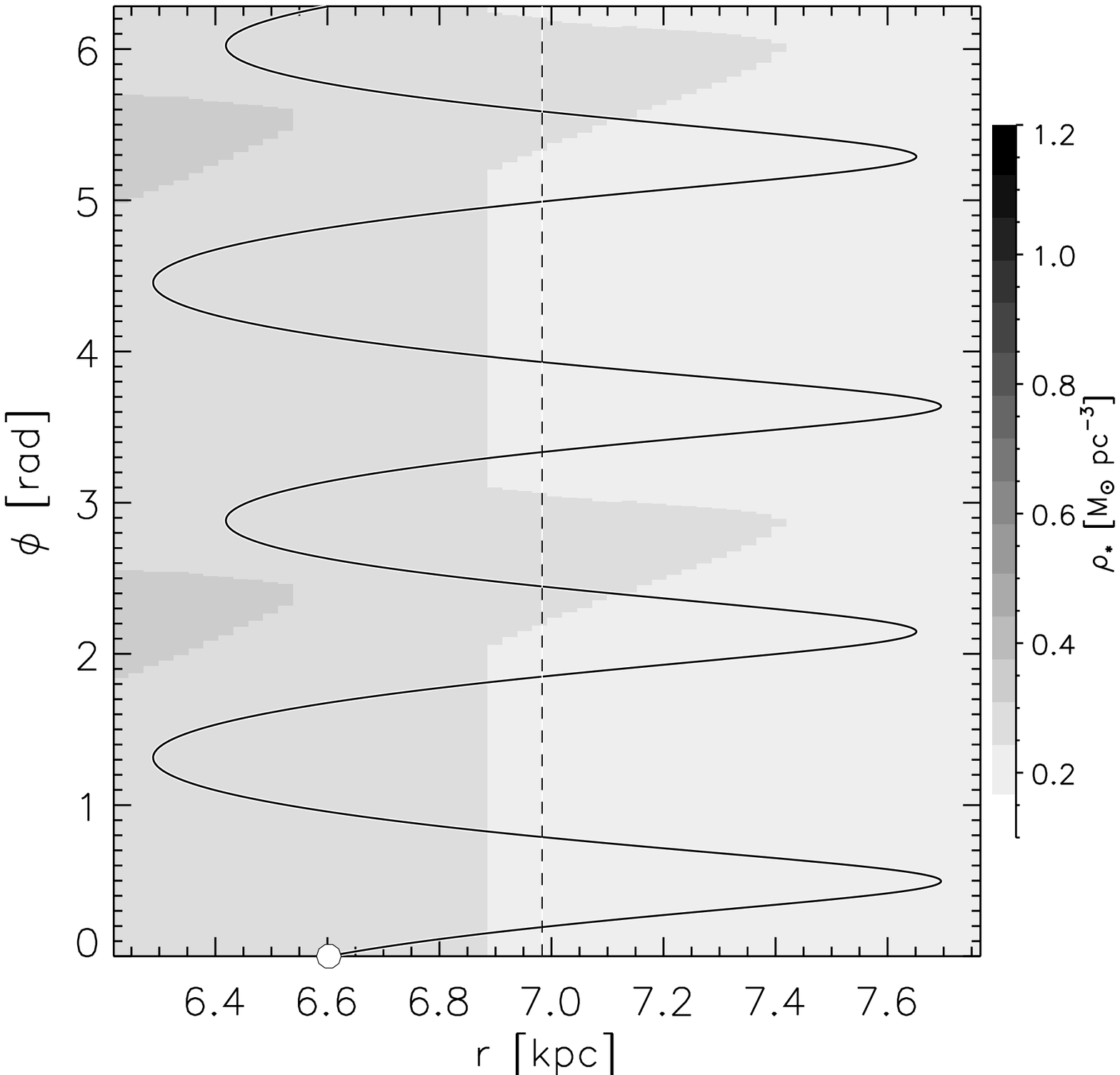}
  \includegraphics[width=0.24\hsize]{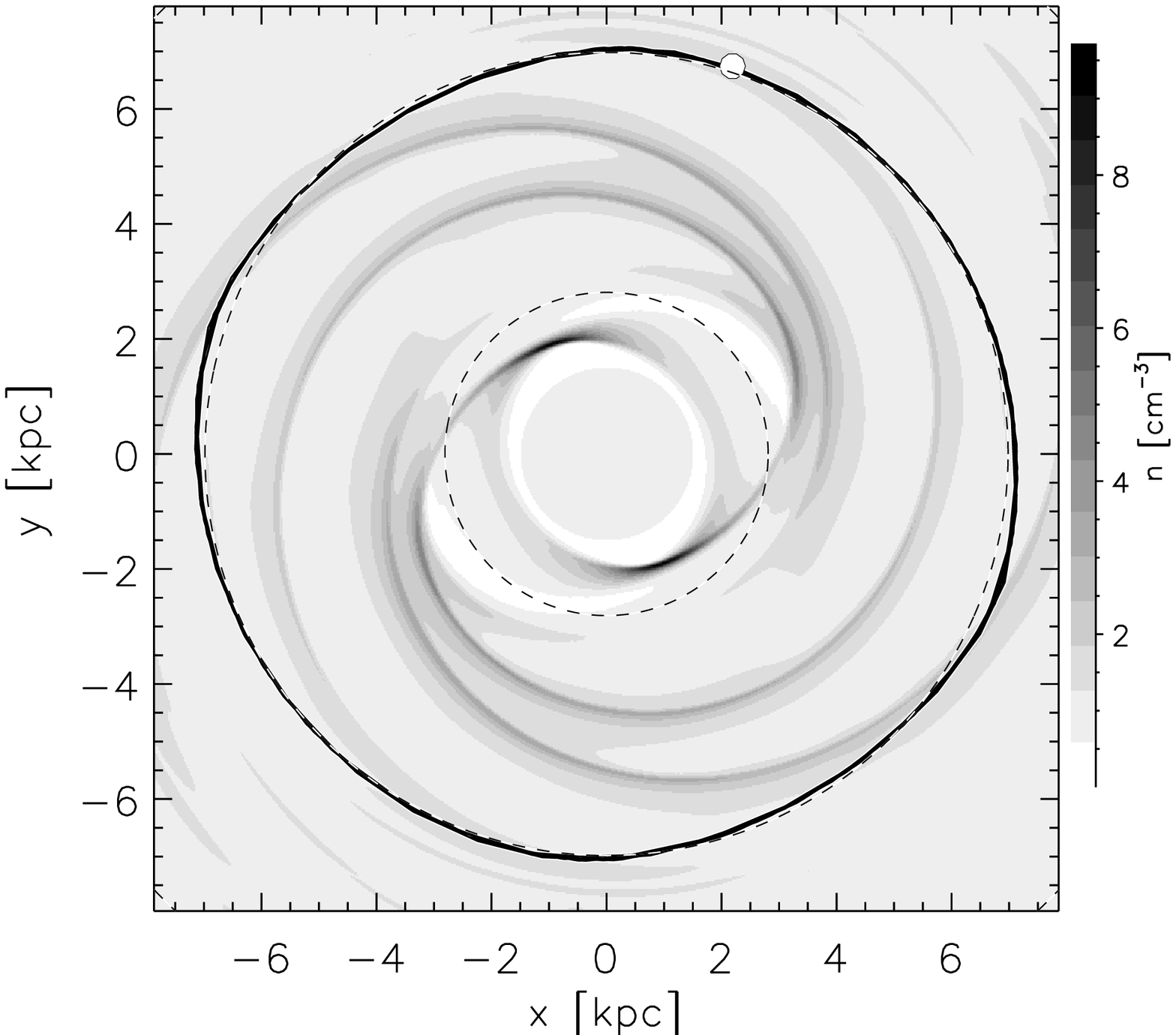}
  \includegraphics[width=0.24\hsize]{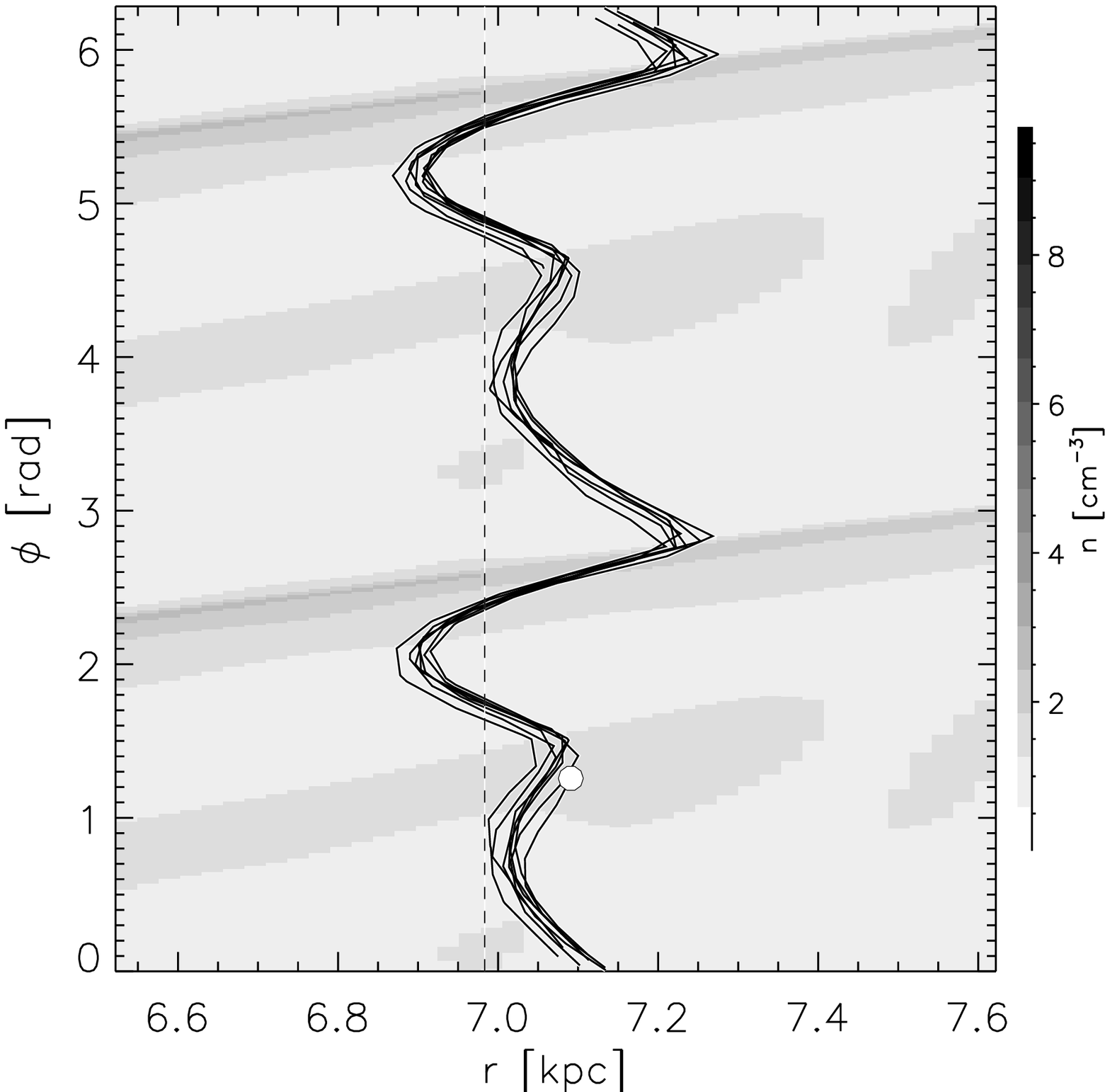}
  \\
  \includegraphics[width=0.24\hsize]{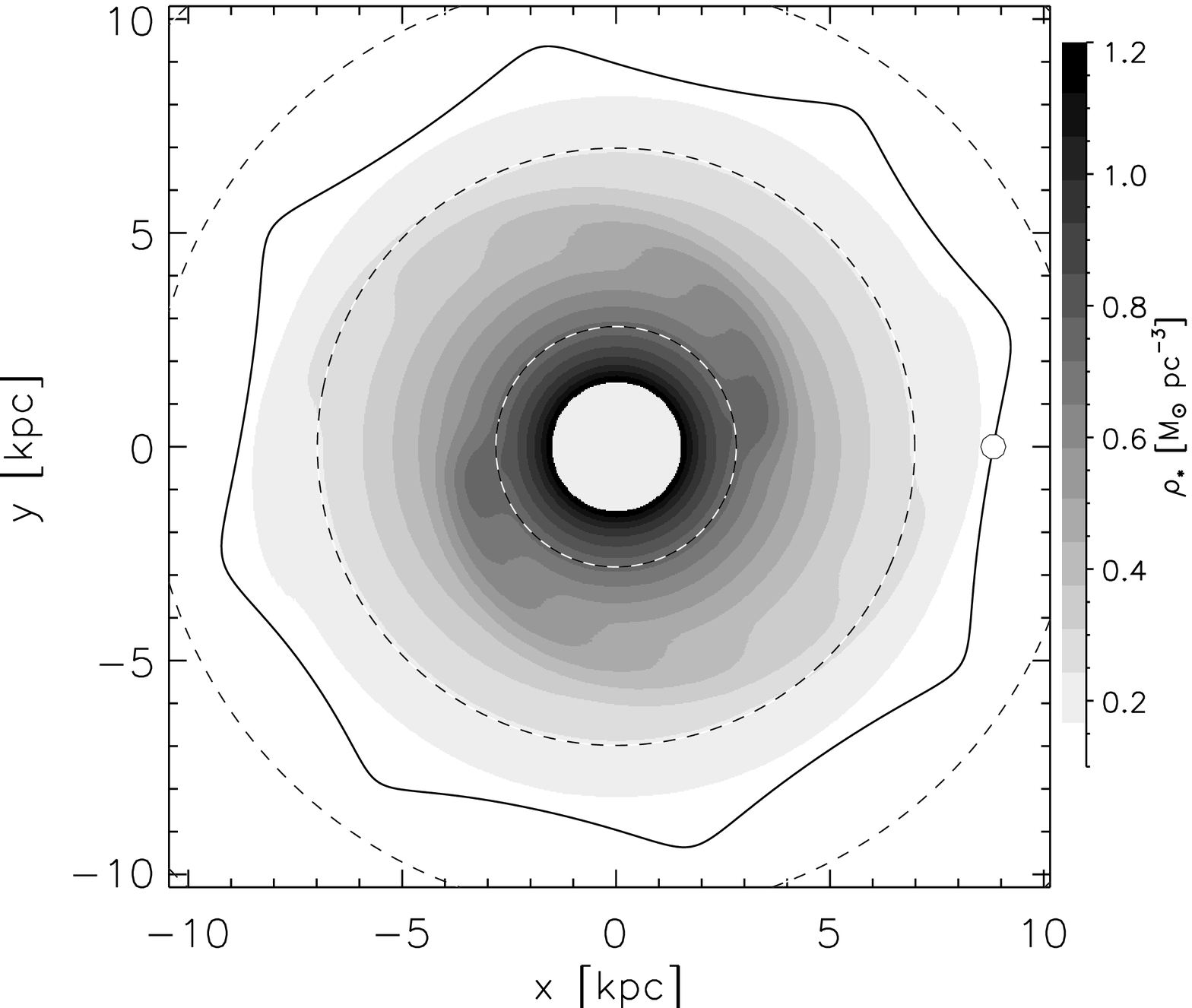}
  \includegraphics[width=0.24\hsize]{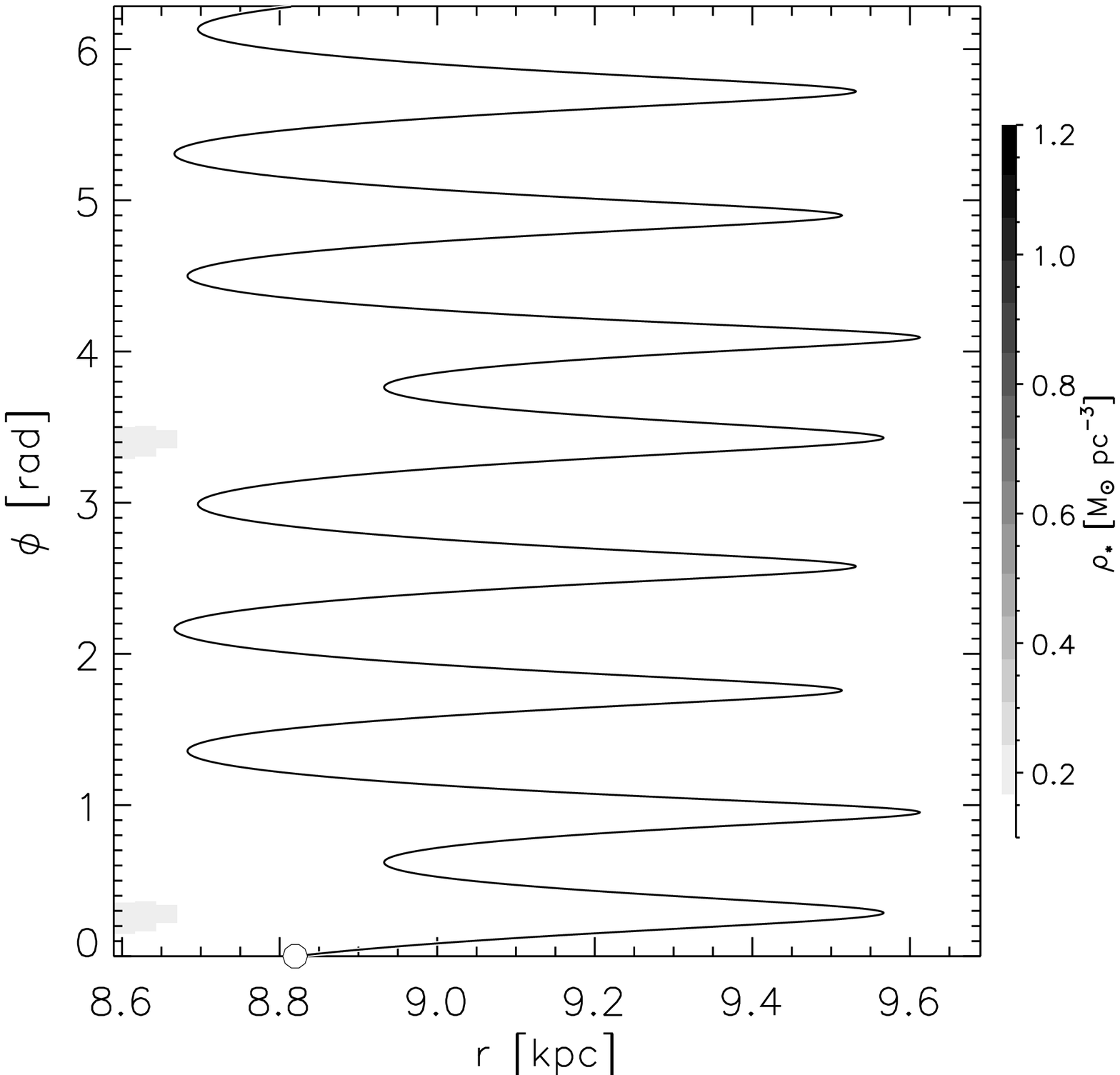}
  \includegraphics[width=0.24\hsize]{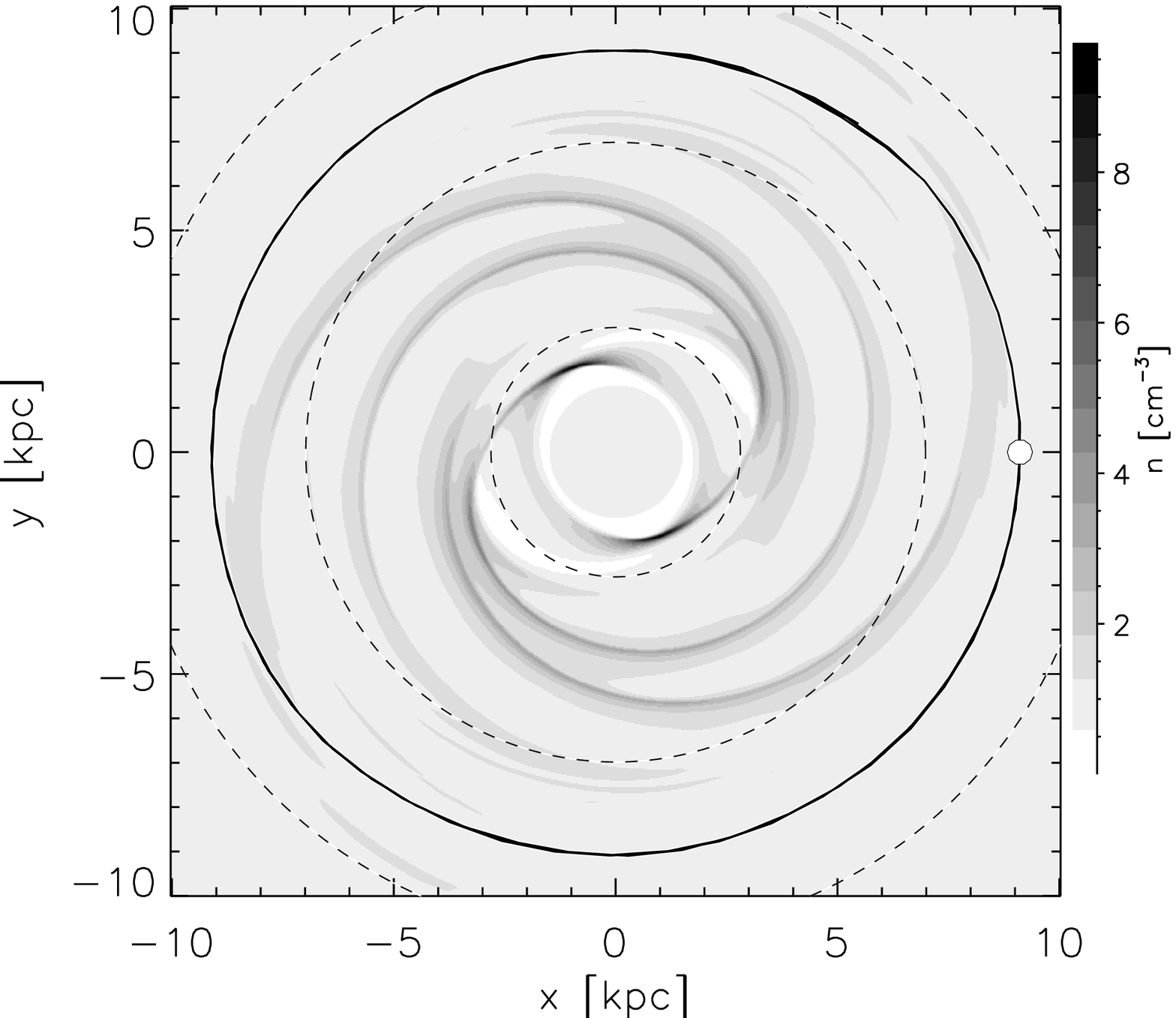}
  \includegraphics[width=0.24\hsize]{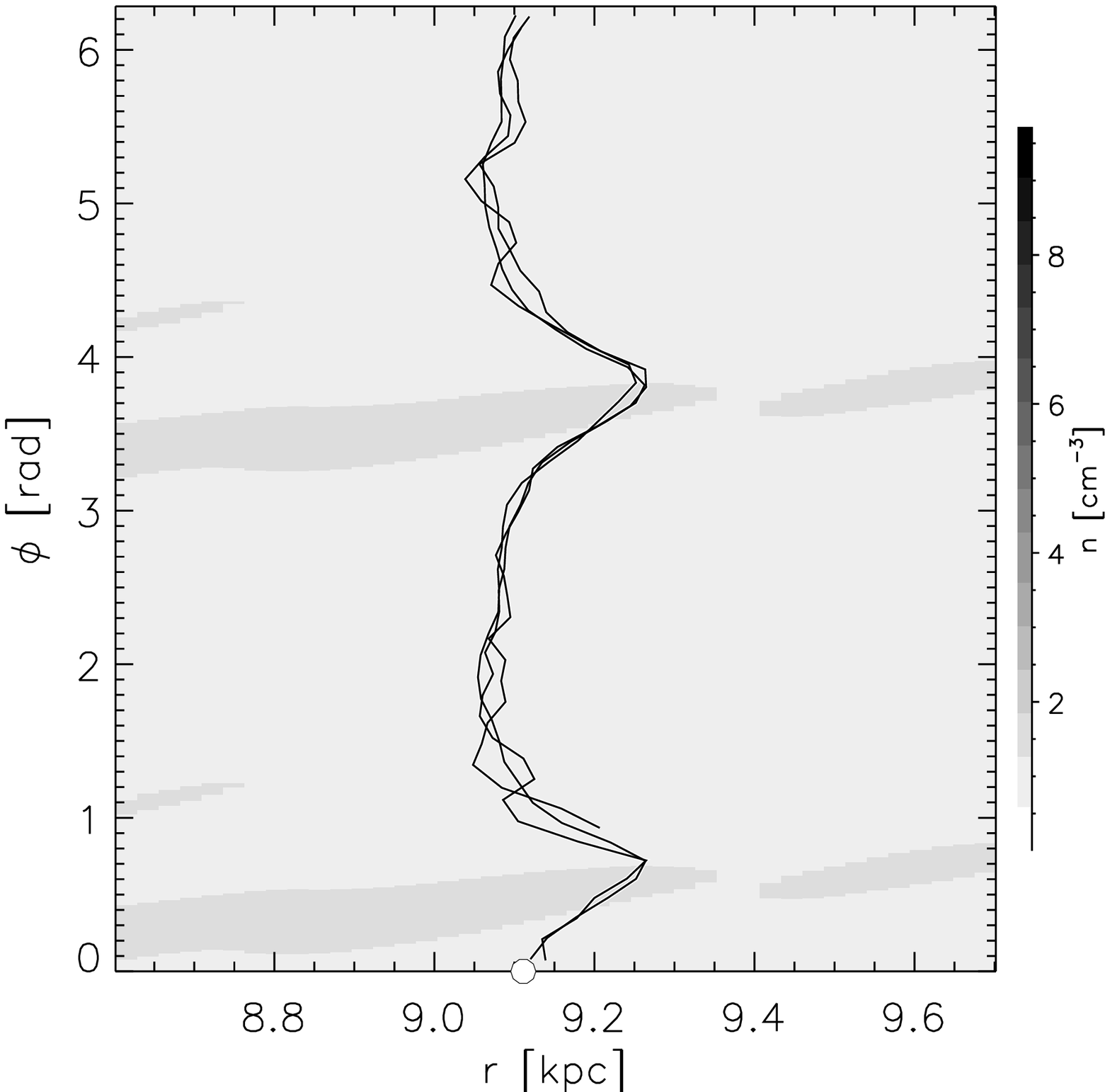}
  \caption{
    Similar to fig.~\ref{fig:orbits_far_res}, for orbits
    near the resonances.
  }
  \label{fig:orbits_near_res}
\end{figure*}

\begin{figure*}
  \begin{center}
  \includegraphics[width=0.30\hsize]{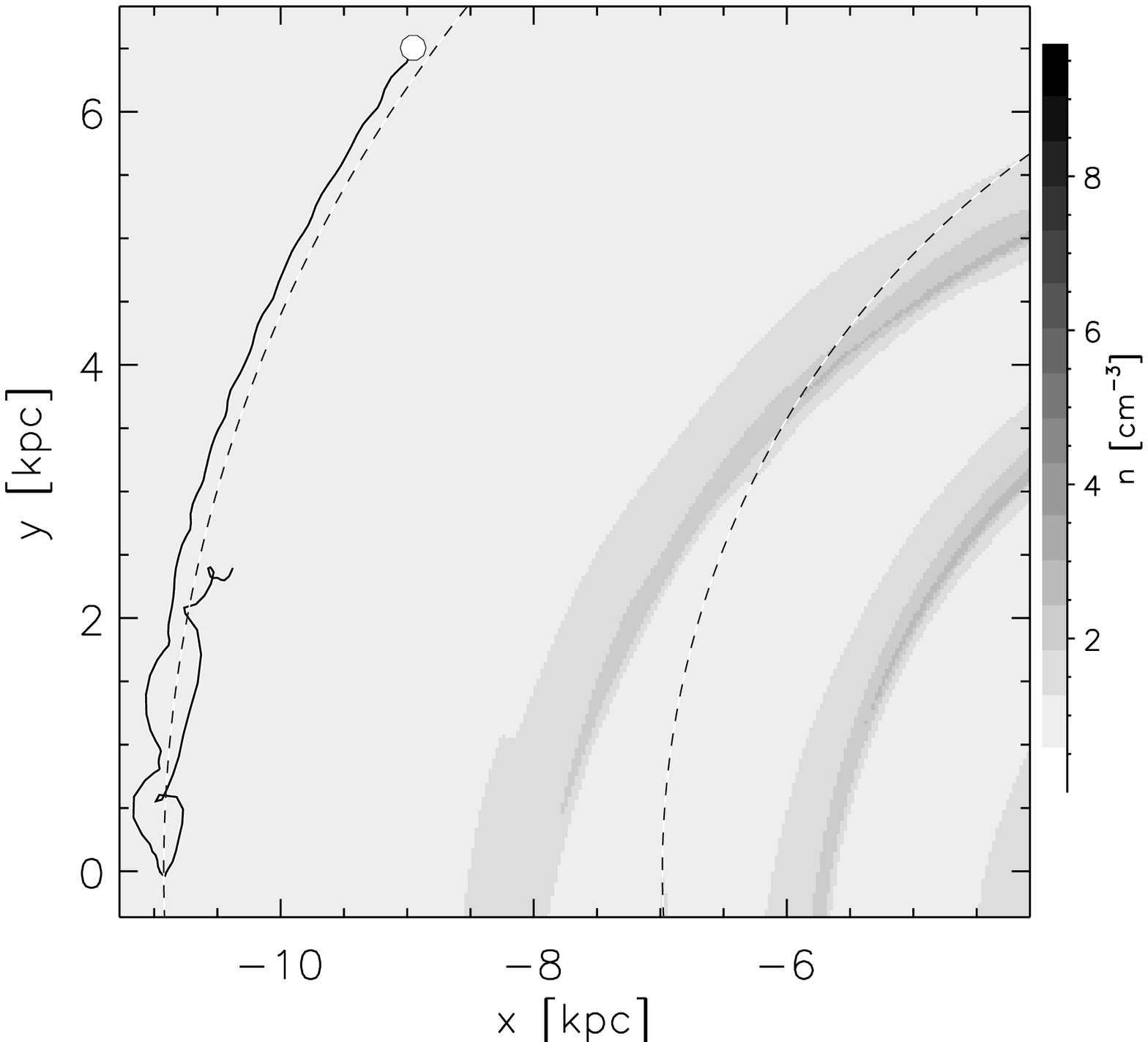}
  \includegraphics[width=0.30\hsize]{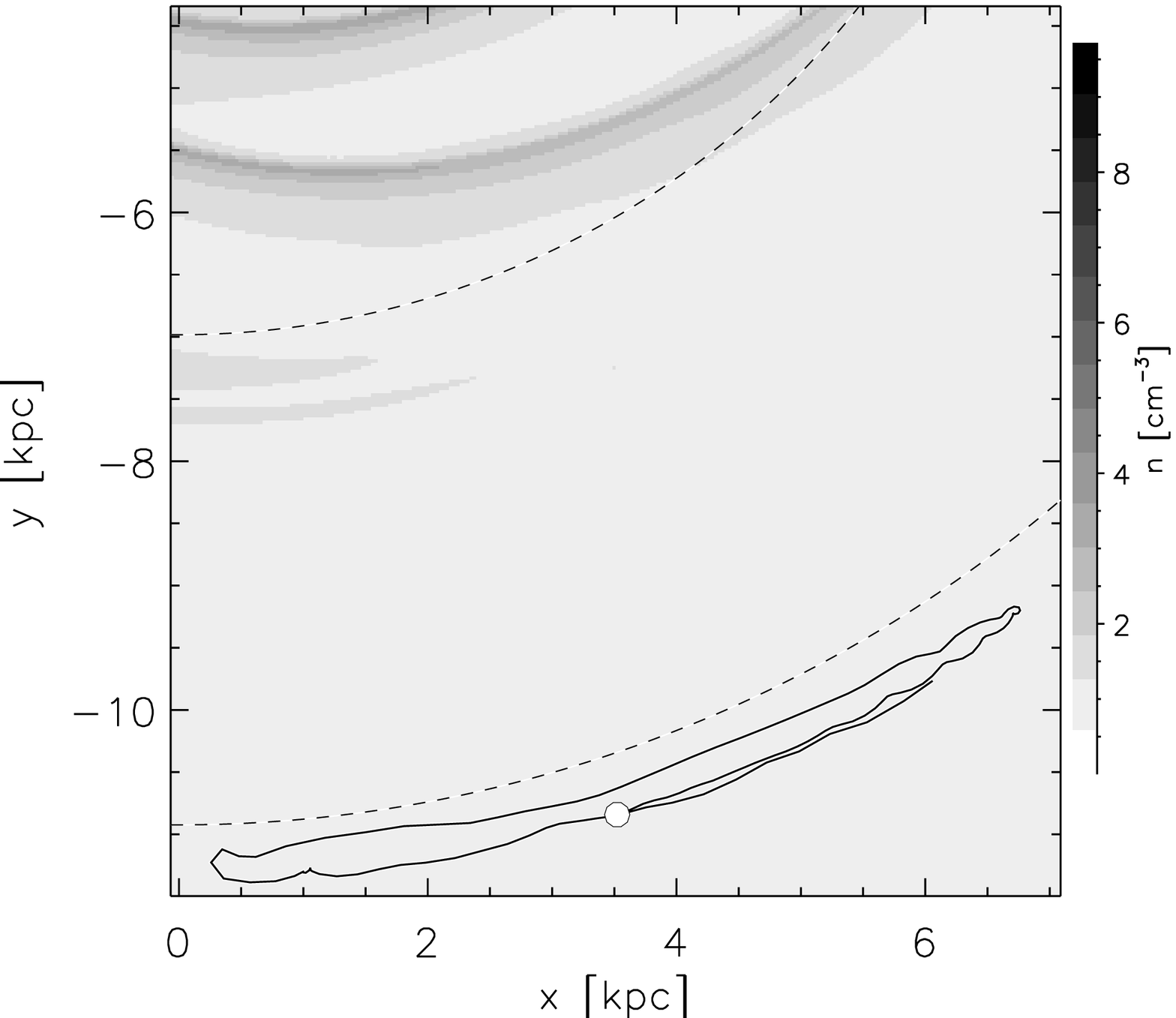}
  \includegraphics[width=0.30\hsize]{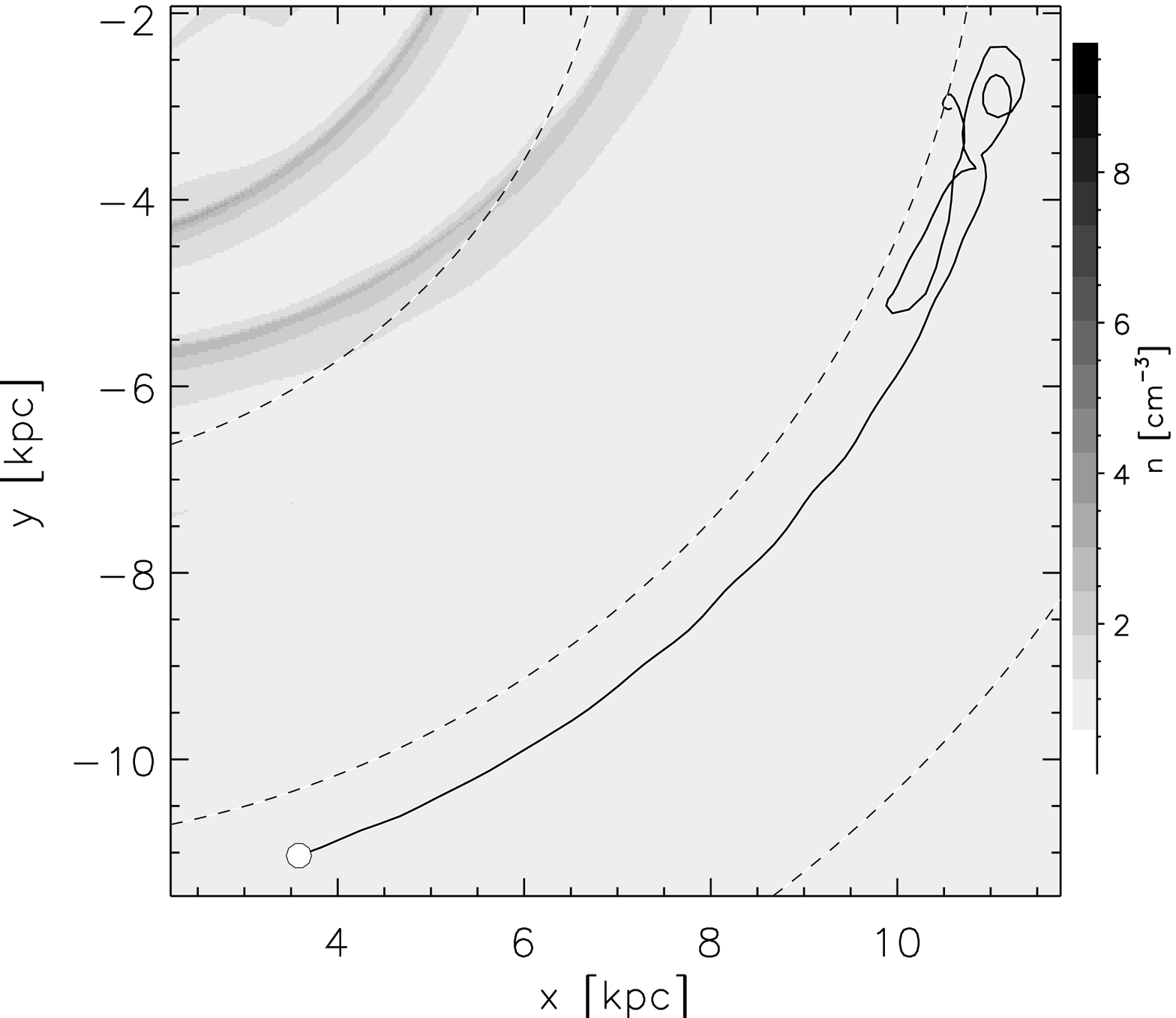}
  \end{center}
  \caption{
    Gaseous orbits near the corotation resonance.
  }
  \label{fig:orbits_near_corot}
\end{figure*}

Although the gas orbits are complex and quite sensitive to the
initial position,
they follow a similar evolution to the stars when regions away from the
ILR are considered.
It is noticeable that the gaseous orbits develop features
that give rise to the second pair of spiral arms (apparent in the
rightmost column of fig.~\ref{fig:orbits_far_res}).
As opposed to the stars, the gas may change direction rapidly by
developing oblique shocks, which generate pressure gradients that
further change the path of the orbit.
Between the ILR and 4:1 resonance, the stellar and gaseous orbits
are similar in the radial range they span, but the shape differs
when the
second pair of gaseous arms becomes stronger, since the shocks at the arm
positions significantly change the path of the gas parcel.
Just outside the 4:1 resonance,
at the same Jacobi constant, we have more than one main
family of periodic orbits coexisting, some of them stable, but the most
of them unstable (see fig.~7 in \citealt{con86}).
Most of stars approach an orbit with a four-fold symmetry;
the gas, on the other hand, selects a twofold symmetric orbit related
to the 2:1 coexisting periodic orbits family,
associated with the only pair of arms
found in the gas in this region.

Near the resonances, the gas and
stars follow quite different orbits (fig.~\ref{fig:orbits_near_res}).
While the stars follow the familiar geometrical shapes, the gas is
unable to follow the large departures from circularity of the
stellar orbits.
While the stars follow oval-shaped periodic orbits near the ILR,
the gas actually tries to avoid that region.
Integrations starting near or inside the ILR decay to the ring/spur
structure just inside $2\kpc$.
This structure is not expected to appear in real galaxies,
since usually it is affected by the
starting position of the spiral arms and should be strongly
influenced by a galactic bar not included in the present model).
In fact, in order to find a gaseous orbit that resembles those
found in the stars near the ILR, we must look for an orbit that
remains outside the ILR region during its time evolution
(the top row in fig.~\ref{fig:orbits_far_res}, for example).
Also, these gaseous orbits present some dispersion, since they are not
quite periodic and actually cross themselves.
This is possible since the simulation does not really reach a steady
state, but instead settles on a periodic cycle, with the arms
oscillating slightly around given positions.

Near the 4:1 resonance, the periodic stellar
orbit follows the familiar rounded-square, while the gas actually
remains close to a circular orbit,
with deviations from circularity of less than $400\pc$
(although these deviations are
systematic and allow for one pair of the gaseous spiral arms to extend
beyond this resonance, as opposed to the stellar pattern).
The strong evolution of the gaseous orbits around the ILR is absent
near the 4:1 resonance.

Further out, the expected banana-shaped orbits appear at some positions,
although the gaseous orbits are greatly distorted due to the turbulence
related to the instability at corotation (fig.~\ref{fig:orbits_near_corot}).
The strong velocity perturbations related to this instability appear to
generate points that act as attractors.
Future work will try to determine whether
these positions are suitable for star formation.
Outside corotation, the gas orbits are nearly circular again,
with deviations less than $0.5\kpc$ from the
initial radius.
This is not surprising since the underlying spiral
perturbation is already weak, representing only $1\%$ of the radial
axisymmetric force at the -4:1 resonance
(the arm potential tapers-off at a radius of $12\kpc$, see
\citealt{pic03}).

\section{Discussion and conclusions}
\label{sec:conclusions}

We performed MHD simulations of a gaseous disk subjected to the
{\tt PERLAS} spiral potential (disregarding the
effect of a galactic bar, which will be explored in the future).
We then studied the actual path the gas follows as it
rotates around the galaxy in order to compare it with the 
stable periodic
stellar orbits existing in that same galactic potential
in order to test the
frequently stated assumption that the gas should follow orbits close to
the periodic stellar orbits that do not cross themselves.
We found that the gas does not always behave that way.

The most obvious difference in the stellar and gas behaviour is that
the gas responds to the two-arm potential with four spiral arms
\citep{shu73,mar04}, organized in two pairs,
each with a tighter pitch angle than the underlying potential.
The four gaseous arms are well defined between the ILR and the 4:1
resonance, more or less straddling the stellar arms
in similar fashion to the secondary compressions described by
\citet{shu73} (see fig.~\ref{fig:densidades}).
Outside the 4:1 resonance, the gaseous arms follow the stellar arms
more closely.
The gas response to stellar arms depend on the details of the
imposed potential (compare, for example, with \citealt{pat97} and
\citealt{vor06}) and will be explored in future work.

As opposed to stars, the gas may develop shocks at the arm
positions, which allow its orbits (understood as the path a gas
parcel follows) to change direction abruptly.
This behaviour generates cusps in the gaseous orbits that are not
present in the stellar orbits.

Setting aside the extra cusps in the gaseous orbits, the stellar and
gaseous orbits are most similar in between the resonances, when
the general shape and radial range are considered.
These are the regions where the periodic stellar orbits are rounder
and the radial excursions are smaller, allowing the gas to more
closely follow the stellar orbits since it will be less influenced
by forces other than gravity.
Near the resonances, however, the stars suffer large radial
excursions that imply regions with small curvature radii so that the
orbit may close on itself (see fig.~\ref{fig:orbits_near_res}).
A gas parcel cannot follow such an orbit\footnote{We have also found
that there is a limit to the ellipticity of a 2:1 orbit (oval ones)
at approximately $0.4$, above which gas cannot settle down readily on
the orbit because of the presence of strong shocks at the apocenter.}
and either looses angular
momentum and moves away from the resonance (such as the orbit near the
ILR), or settles in an orbit with only small radial excursions.
The gravitational forcing that maintains the resonance is still
present;
so, even as the gas follows a nearly circular orbit, its
velocity is not uniform, thus generating pressure waves that allow
the spiral arm to extend across the resonant radius;
this is apparent, for example, across the 4:1 resonance.

Although banana-type orbits should be present at the corotation
radius, the gas only develops such orbits just beyond that
resonance.
Generally speaking, the gas experiences changes always in the outer
side of the resonant radii, for example, the second pair of arms start
and end just outside the ILR and 4:1 resonance, respectively.
This particular element will be further explored in future work.

When stating that the gas should follow stable periodic stellar
orbits that
do not cross, a key assumption is that the gas falls into a steady
state.
Small oscillations of the gaseous arms (an effect exacerbated
when the vertical structure of the gaseous disk is considered,
see \citealt{gom04}) and magnetohydrodynamic instabilities (like that
present at corotation, see \citealt{mar13}), distort the gaseous orbits
so that they may cross themselves.
Other phenomena not considered here, like star formation and
feedback, will further disrupt the path a gas parcel should follow.

On the other hand, in chaotic regions, such as the corotation
zone, the phase space for stable periodic orbits is severely
reduced. Gas dynamics behaviour in this region can not be
attributable to stable simple paths given by periodic orbits
\citep{cha11}. In these regions, chaotic stellar orbits
might be the ones supporting large scale structures, and gas might
not be able to follow chaotic orbits since their trajectories are, in
general, self-intersecting.

Since observations of the spiral structure of the Milky Way are well
fitted by four arms in the gas and two in the stars, the orbits that
support such structures cannot be the same.
The differences in the orbital behaviour of gas and stars shown here
highlights the difficulty of extrapolating the results of
stellar dynamics to gas dynamics.
Even if the motion is dominated by gravity, other physical processes
may have a strong impact on the overall behaviour of a galactic disk.

\section*{Acknowledgments}

The authors wish to thank P. Patsis for useful discussions on the
subject, W. Henney for reviewing the manuscript, and an anonymous
referee for comments that led to a much improved paper.
This work has received financial support from UNAM-DGAPA PAPIIT grants
IN106511 to G.C.G. and IN110711 to B.P.


\label{lastpage} 
\end{document}